\documentclass[aps,prl,twocolumn,notitlepage,showpacs,preprintnumbers,amsmath,amssymb,revsymb,doi=false,isbn=false,url=false]{revtex4-1}
\usepackage{graphicx}
\usepackage{color} 
\usepackage{txfonts}
\usepackage{dcolumn}
\usepackage{bm}
\usepackage{braket}
\usepackage[colorlinks=true, urlcolor=blue, citecolor=blue,
linkcolor=red]{hyperref}

\definecolor{gray}{rgb}{0.5,0.5,0.5}

\def\braket#1{\mathinner{\langle{#1}\rangle}}

\def\mychi{\raisebox{0.35ex}{$\chi$}}

\begin{document}

\title{Superconducting grid-bus surface code architecture for
  hole-spin qubits}

\author{Simon E. Nigg$^1$}\email[Corresponding author:
]{simon.nigg@unibas.ch}
\author{Andreas Fuhrer$^2$}
\author{Daniel Loss$^1$}
\affiliation{$^1$Department of Physics, University of Basel,
  Klingelbergstrasse 82, 4056 Basel, Switzerland}
\affiliation{$^2$IBM Research - Zurich S{\"a}umerstrasse 4, 8803
  R{\"u}schlikon, Switzerland}
\date{\today}

\begin{abstract}
We present a scalable hybrid architecture for the 2D surface
code combining superconducting resonators and hole-spin qubits in
nanowires with tunable
direct Rashba spin-orbit coupling. The back-bone
of this architecture is a square lattice of capacitively coupled
coplanar waveguide resonators each of which hosts a nanowire
hole-spin qubit. Both the frequency of the qubits and their coupling to
the microwave field are
tunable by a static electric field applied via the resonator center pin. In the dispersive
regime, an entangling two-qubit gate can be realized via a third
order process, whereby a virtual photon in one
resonator is created
by a first qubit, coherently transferred to a neighboring resonator, and
absorbed by a second qubit in that resonator. Numerical simulations
with state-of-the-art coherence times yield gate fidelities
approaching the $99\%$
fault tolerance threshold.
\end{abstract}

\maketitle

Scalability is central to the ongoing efforts
towards fault tolerant quantum computation~\cite{Taylor2005,
  Chen-2014, Nemoto-2014, Hill-2015, Billangeon-2015, Brecht2016, Karzig-2016-arxiv}. Owing to its high error
rate threshold and its benign requirement
of only local qubit interactions, the surface
code~\cite{Bravyi-1998} is a promising candidate to achieve fault
tolerance. Superconducting circuits, with their long coherence times
and high-level of controlability, have emerged as an ideal
platform for a physical implementation of the surface
code~\cite{Helmer-2009, DiVincenzo-2009, Fowler-2012a, Barends2014,
  Corcoles2015, Kelly2015, Takita-2016}.
At the heart of this approach lies the coherent light-matter
interaction between the electric dipole moment of a superconducting condensate and quantized
microwave fields~\cite{Blais-2004a}. This interaction
 however is a double-edged sword. On the upside, it enables the readout
 and control
 of
superconducting qubits and of their interaction with each other via the quantum
bus~\cite{Blais-2007a, Majer-2007a}. On the
downside, the presence of an electric dipole moment means that un-monitored degrees
of freedom, such as thermal and quantum fluctuations of the field, couple to
the qubits and limit their
coherence~\cite{Gambetta-2006}. Moreover, in a multi-qubit
system, the accumulation of errors due to off-resonant couplings
represents a serious problem for scalability~\cite{Corcoles2015,
  Kelly2015, Blumoff-2016, Takita-2016}. The ability to tune the
light-matter coupling on and off on-demand is thus highly
desirable. Superconducting qubits with tunable
qubit-resonator coupling have been realized~\cite{Gambetta-2011,
  Srinivasan-2011, Eichler-2015, Gengyan-2016-arxiv}, but their robustness is limited
since they rely on quantum coherent
interference at a symmetry point.
\begin{figure}[ht]
\includegraphics[width=\columnwidth]{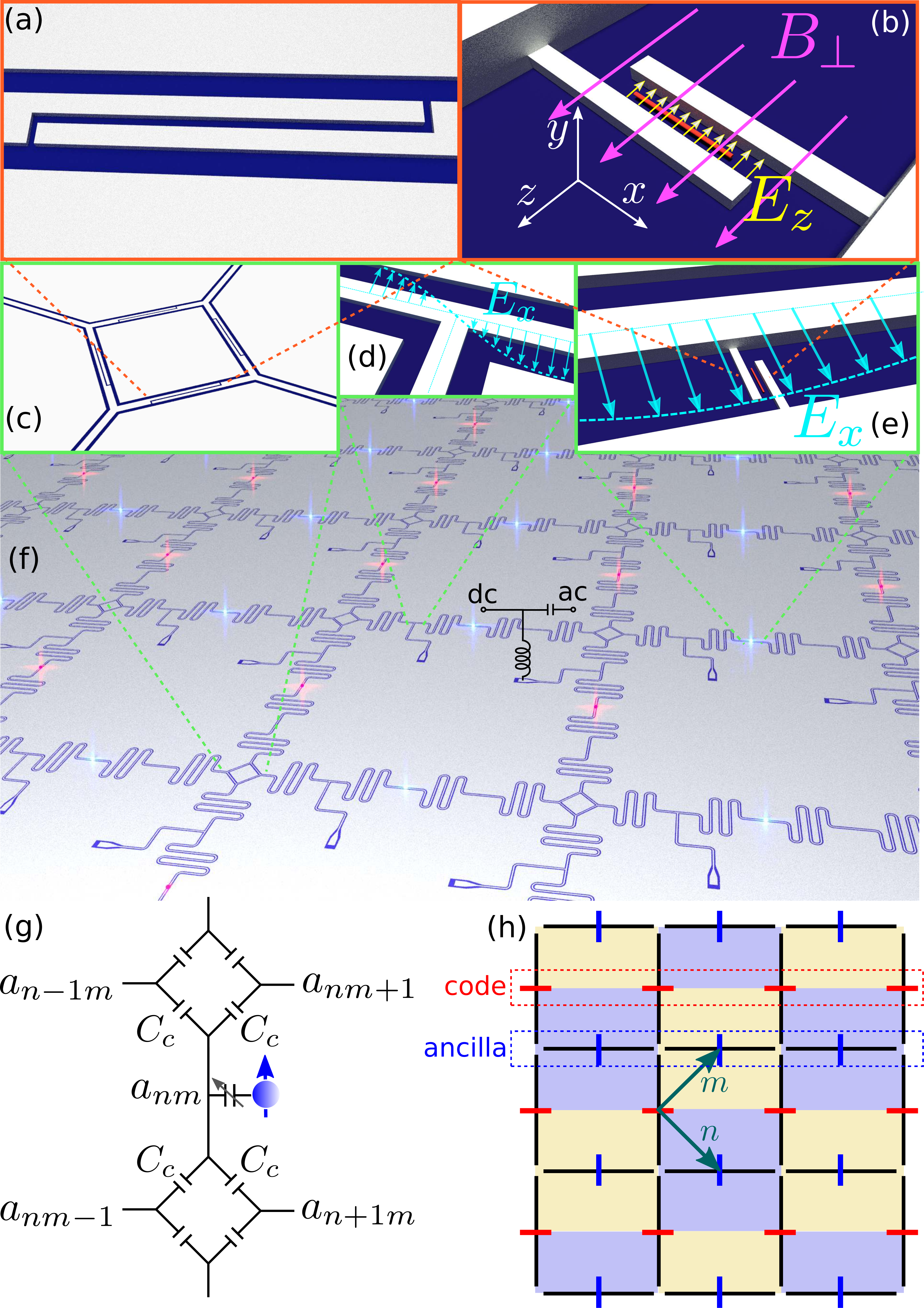}\caption{Grid-bus
  surface code architecture. {\bf (a)} and {\bf (c)}: Four-way capacitor design
  minimizing undesired cross-couplings. {\bf (b)} and {\bf (e)}: A nanowire
  hole-spin qubit inside a capacitor in the trench of the
  resonator. The electric field perpendicular to the wire $E_z$ is controlled
  by voltage biasing the center conductor via the bias-tee shown in
  (f). {\bf (d)}: Resonator drive port placed at a node of
  the ac field $E_x$. {\bf (f)}: Resonator grid layout. The light gray areas
  represent the superconductor thin film on top of the dielectric
  substrate (dark blue). The red and blue dots at the center of each resonator
  indicate the positions of the nanowire qubits. {\bf (g)}: Each
  resonator couples to four neighboring resonators. {\bf (h)}: Resonator (black
  lines) and qubits (red and blue bars) arranged in a square
  lattice. The red bars denote
  code qubits while the blue bars denote ancilla qubits. The colored rectangles represent the two types of plaquettes of
  the surface code (e.g. $XXXX$ or $ZZZZ$). The basis vectors on the lattice are indicated by dark
  green arrows labeled $n$ and $m$.\label{fig:setup}}
\end{figure}

The recent discovery by~\citet{Kloeffel-2011} of an electrically induced spin-orbit
interaction of Rashba type in the low energy hole states of Ge/Si
(core/shell) nanowires
provides an attractive alternative to realize a tunable coupling qubit. In this case the qubit is encoded in two orthogonal
dressed spin states of a hole confined in a nanowire quantum
dot. Hole spins are particularly attractive since their p-wave orbitals have minimal
overlap with the nuclei resulting in long
coherence times~\cite{Hu2012, Kloeffel-2013, Prechtel2016,
  Watzinger-2016} and have recently been demonstrated to be compatible
with industrial CMOS technology~\cite{Maurand-2016-arxiv}. Crucially
the strong direct Rashba spin-orbit interaction (DRSOI) is controlled by
an {\em external} electric field applied perpendicular to the
wire~\cite{Kloeffel-2011, Kloeffel-2013}. This
enables the electrostatic control of the coupling between the spin
degree of freedom and the electromagnetic field along the wire.

In this letter, we propose a scalable surface code architecture
obtained by combining nanowire hole-spin qubits with a novel coplanar
waveguide resonator grid structure. The latter can be viewed as a generalization of the
celebrated 1D quantum bus architecture~\cite{Blais-2007a, Majer-2007a}
to two dimensions. Furthermore, owing
to the small size of the nanowire qubits, a few tens of nanometers in length, they can
be entirely embedded within the microwave resonators allowing for more
compact resonator geometries with enhanced vacuum field strengths. The electrostatic
fields required to tune the microwave-qubit coupling, are
provided {\em in-situ} by voltage biasing the
resonator center conductor thus reducing the number of
required leads.

The system we consider is depicted schematically in
Fig.~\ref{fig:setup}. It consists of a square lattice of coplanar
microwave resonators, with a nanowire qubit
placed at the field anti-node of each resonator. Here we consider
full-wave resonators where the
resonator length equals the wavelength $\lambda$ and the qubits are placed at the
central anti-node. Each resonator is capacitively coupled to
four neighboring resonators forming a horizontal ``H'' shape as shown in
Fig.~\ref{fig:setup}~(g). The nanowires, each containing a single
spin-orbit qubit, are situated inside the
trenches between the center conductor and the ground plane defining the
resonator, as depcited in insets (b) and (e) of 
Fig.~\ref{fig:setup}. The qubit is thus fully embedded within the resonator.
The electromagnetic
fields are only weakly screened inside the semiconductor of the
nanowire enabling a strong coupling between the qubit
and the ac field component along the wire~\cite{Kloeffel-2013}.

To characterize this system, we start by considering an isolated site of the lattice composed of one
resonator and one hole-spin nanowire qubit. The nanowire is oriented along the
$x$-axis and a magnetic field is applied along $z$. We describe the
hole harmonically confined along the wire by the 1D effective Hamiltonian~\cite{Maier-2014}
\begin{align}\label{eq:6}
\bm H_h=\frac{\bm p^2}{2m}+\frac{1}{2}m\omega_h^2\bm x^2+\alpha_{\rm
  DR}\bm\sigma^y\bm p+\frac{g\mu_{\rm B}B_{\perp}}{2}\bm\sigma^z.
\end{align}
Here $\alpha_{\rm DR}$ is the strength of the DRSOI and $B_{\perp}$ denotes
the magnetic field strength perpendicular to the axis of the wire. The hole
furthermore couples to the electromagnetic field of the
resonator and this is described in dipole approximation via
the Hamiltonian
\begin{align}
\bm H_{c}=eE_{\rm rms}\bm x(\bm a+\bm a^{\dagger})+\hbar\omega_r\bm
  a^{\dagger}\bm a.
\end{align}
Here $E_{\rm rms}=\frac{1}{W}\sqrt{\frac{\hbar\omega_r}{cl}}$ is the
$x$-component of the anti-node vacuum root mean square field of the CPW resonator with resonance
frequency $\omega_r$, trench width
$W$, length $l$ and capacitance per unit length $c$.
The full Hamiltonian is $\bm H=\bm H_h+\bm H_{r}$. The effect of the
spin-orbit coupling is seen most clearly upon performing the unitary
transformation $\bm U=\exp\left[ i(\bm x/\ell_{\rm SO})
  \bm\sigma^y\right]$, where the spin-orbit length
$\ell_{\rm SO}=\hbar/(m\alpha_{\rm DR})$ characterizes the length over
which the spin flips due to spin-orbit coupling in the absence of a
magnetic field. This generalizes the
semiclassical approach of~\cite{Flindt-2006, Trif-2007} to the quantum regime. In the limit where $\ell_{\rm SO}\gg
x_{\rm ZPF}=\sqrt{\hbar/(2m\omega_h)}$, the mixing of orbital and spin
degrees of freedom is weak and the transformed Hamiltonian reads~\cite{supp_mat}
\begin{align}
\bm{H}\simeq\bm H_{c}+\frac{\bm
  p^2}{2m}+\frac{1}{2}m\omega_h^2\bm x^2+\frac{\hbar\omega_Z}{2}\left(
  \bm\sigma^z-\frac{2\bm x}{\ell_{\rm SO}}\bm\sigma^x \right).
\end{align}
Here we have suppressed a c-number term and defined the
Zeeman frequency
$\omega_Z=g\mu_{\rm B} B_{\perp}/\hbar$.

We are interested in the regime
where $\omega_h\gg\omega_r,\omega_Z$ such that the hole remains in its ground
state. In
this regime, we can adiabatically eliminate the center of mass motion
of the hole~\cite{supp_mat}. The dynamics of the hole-spin coupled to the resonator is
then captured by an effective Jaynes-Cummings model
\begin{align}\label{eq:7}
\frac{\bm H_{\rm JC}}{\hbar}=\omega_r\bm
a^{\dagger}\bm a+\frac{\omega_Z'}{2}\bm \sigma^z +\nu\left( \bm
  a\bm\sigma^++\bm a^{\dagger}\bm \sigma^- \right),
\end{align}
where the transition frequency of the qubit is determined by the
renormalized Zeeman splitting
\begin{align}\label{eq:2}
\omega_Z'=\omega_Z\left[ 
  1-\frac{\omega_Z}{\omega_h-\omega_Z}\left( \frac{x_{\rm
  ZPF}}{\ell_{\rm SO}} \right)^2  \right],
\end{align}
and the spin-field coupling strength is given by
\begin{align}\label{eq:3}
\nu=\frac{\beta\omega_Z}{\omega_h-\omega_r}\left( \frac{x_{\rm
  ZPF}}{\ell_{\rm SO}} \right).
\end{align}
Here $\beta = eE_{\rm rms}x_{\rm ZPF}$ is the dipole coupling
strength between the hole in the motional ground state and the vacuum
field of the resonator. Importantly $\ell_{\rm SO}$ depends,
via the spin-orbit coupling strength $\alpha_{\rm DR}$, on the electric field
component $E_z$ perpendicular to the wire. In the weak field limit
$\alpha_{\rm DR}\propto E_z$ and so the coupling $\nu$ increases linearly with
$E_z$ while $\omega_Z'$ decreases quadratically. Thus the
``off'' state, $E_z=0$, corresponds to a sweet-spot for the qubit where
it is protected against
fluctuations of the electric field to linear order. A non-perturbative treatment, of which the
above expressions (\ref{eq:2}) and~(\ref{eq:3}) are the leading order expressions, can be found
in~\cite{Kloeffel-2013}. 
For Ge/Si nanowires, the Zeeman splitting in (\ref{eq:2}) reaches
the GHz frequency regime for magnetic field strengths around one
hundred milli Tesla. 
We emphasize that our architecture is compatible with a magnetic field
parallel to the plane of the superconducting resonator, mitigating
adverse effects on the resonator quality factor~\cite{Samkharadze-2016}.
A
strength of our architecture is that the required electrostatic control field can be
generated without the need for additional bias lines. This is achieved by applying a voltage bias between the center conductor
and the ground plate of the resonator via a low-pass filtered T-junction contact formed
at a field node as depicted in inset (d) of
Fig.~\ref{fig:setup}. By using a bias-tee, the same port can
be used to drive the resonator. When the qubit-resonator coupling is on,
this enables fast qubit rotations around any axis in the $x-y$ plane
of the Bloch sphere.

Scaling up, we next consider
an $N\times M$ lattice of such resonators, where each resonator is
coupled capacitively to four neighboring resonators as illustrated in
Fig.~\ref{fig:setup}~(g). A global in-plane magnetic field is applied
at an angle with the nanowires (ideally $\pi/4$ for equal strength
coupling to nanowires in both orientations). Because of the
strong suppression of the g-factor along the axis of the
wires~\cite{Maier-2013, Brauns-2016}, we
consider for each wire only the perpendicular component of the
magnetic field justifying the
applicability of Eq.~(\ref{eq:7}) also in this case. In the rotating wave
approximation, the dynamics on the lattice can be modeled by
the Jaynes-Cummings-Hubbard Hamiltonian~\cite{Guanyu-2013}
\begin{align}\label{eq:1}
&\frac{\bm H}{\hbar} =\sum_{n=1}^N\sum_{m=1}^M\left[ 
  \frac{\omega_Z}{2}\bm\sigma_{nm}^z+\omega_r\bm
  a_{nm}^{\dagger}\bm a_{nm}^{}+\frac{\nu_{nm}}{2}\left( \bm
  a_{nm}^{}\bm\sigma_{nm}^++{\rm h.c.} \right) \right]\\
&+J\sum_{n,m}\left( \bm a_{nm}^{\dagger}\bm a_{nm+1}^{}+ \bm
  a_{nm}^{\dagger}\bm a_{nm-1}^{}+ \bm a_{nm}^{\dagger}\bm
  a_{n+1m}^{}+ \bm a_{nm}^{\dagger}\bm a_{n-1m}^{}+{\rm h.c.}\right).\nonumber
\end{align}
The inter-resonator coupling strength is given by
$J=2\hbar\omega_r\frac{C_c}{C+4C_c}$, in terms of the mode
frequency $\omega_r$, the coupling
capacitance $C_c$ and the effective self-capacitance of the resonator
mode $C=cl$. The tunable
spin-resonator coupling of lattice site $(n,m)$ is denoted with $\nu_{nm}$.

A scalable implementation of the surface code requires:
\begin{itemize}
\setlength\itemsep{0.0em}
\item[(i)] Two-qubit gates between nearest neighbors on a lattice.
\item[(ii)] Arbitrary single-qubit rotations.
\item[(iii)] Individual qubit readout in the computational basis.
\item[(iv)] Parallelizability.
\end{itemize}
Conditions (i) and (ii) together allow one to encode the error
syndrome onto ancilla qubits and (iii) allows one to read out the
error syndrome. Condition (iv) means that the gates
must be performed in parallel so that the time for a single syndrome measurement
cycle does not increase with the lattice size. In theory all stabilizer operator measurements
could be done simultaneously, since per definition the stabilizer operators commute
with each other. However, in practice when the
measurements of multi-qubit stabilizer operators are decomposed into sequences of single
and two-qubit gates between pairs of qubits, a certain degree of sequentiality is
unavoidable. In the following we show how our architecture meets the requirements
(i) to (iv).

{\em Single-qubit gates}. To address a particular qubit, the center
conductor of the corresponding resonator is voltage biased, generating
an electric field $E_z=E_z^*$ perpendicular to the wire (see
Fig.~\ref{fig:setup} (b)). This effectively turns on the
DRSOI and couples the
qubit to the ac field. Single-qubit rotations around any axis in
the $x-y$ plane of the Bloch-sphere can then be performed in a standard
way~\cite{Blais-2004a}, by driving the resonator mode at the Lamb and
Stark shifted qubit resonance
frequency with a coherent microwave drive of appropriate
phase (see Fig.~\ref{fig:gate_fids} (a) and~\cite{supp_mat}). By concatenating rotations around different axes, arbitrary
rotations on the Bloch sphere can be
generated. Interestingly, the electric tunability of the Zeeman
splitting provides a shortcut for single-qubit phase gates: To acquire a phase $\theta$ one simply
has to bias the center conductor for a duration
$T=\theta/\Delta\omega_Z$ with
$\Delta\omega_Z=\omega_Z(E_z=0)-\omega_Z(E_z=E_z^*)$. Because each
resonator is coupled only to its four orthogonal neighboring resonators,
single-qubit gates can be performed in parallel on all code qubits and
separately on all ancilla qubits by alternatingly coupling one set of
qubits to the grid-bus while the other remains uncoupled.

\begin{figure}[ht]
\includegraphics[width=\columnwidth]{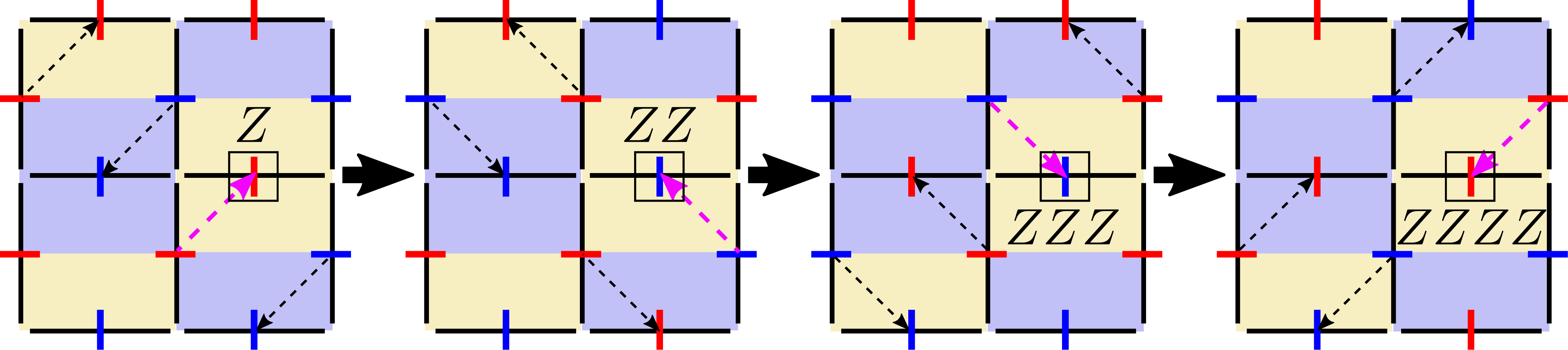}\caption{Frequency
  layout for parallelization of syndrome mapping in four steps. Red (blue) bars
  denote qubits at frequency $\omega_Z^{\rm (r)}$ ($\omega_Z^{\rm (b)}$). The dashed black
arrows indicate which couplings are resonant, i.e. active in a given
configuration. The mapping of a $ZZZZ$ stabilizer is highlighted as
an example (magenta arrows).\label{fig:parallel}}
\end{figure}

\begin{figure*}[ht]
\includegraphics[width=\textwidth]{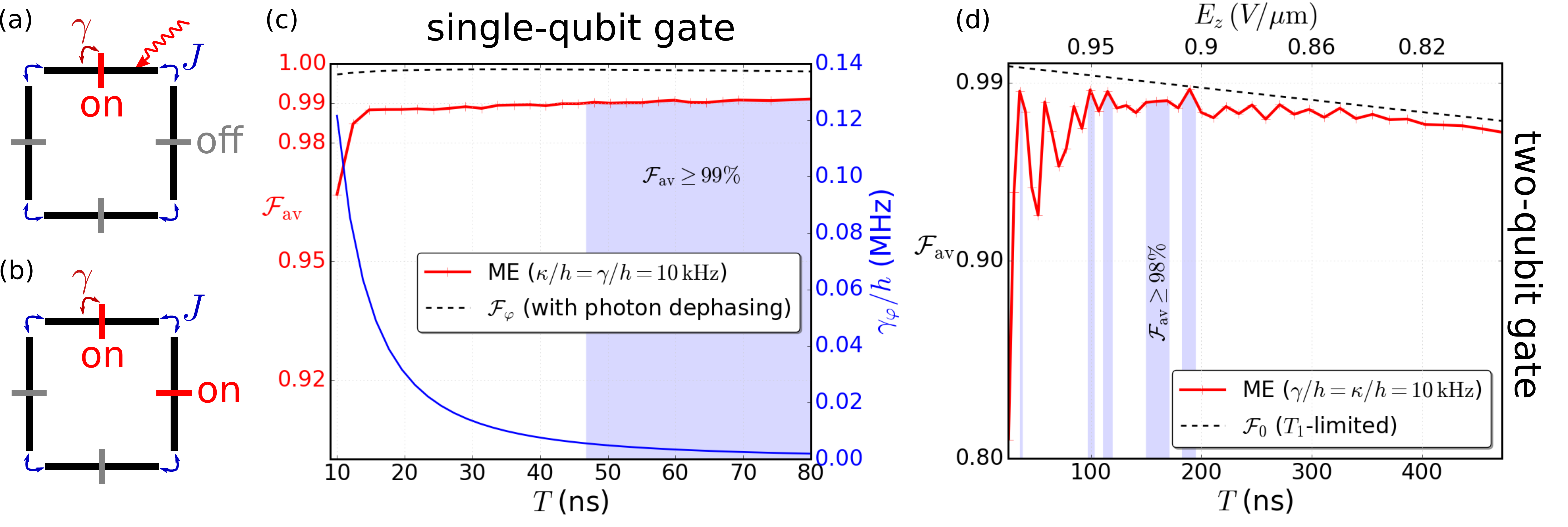}\caption{Gate
  fidelity averaged over initial states on the Bloch sphere: $\mathcal{F}_{\rm
  av}=\frac{1}{4\pi}\int_{0}^{2\pi}{\rm d}\varphi\int_{0}^{\pi}{\rm
  d}\theta\sin(\theta)\mathcal{F}(\theta,\varphi)$, with
$\mathcal{F}(\theta,\varphi)=\braket{\theta,\varphi|\bm\rho(T)|\theta,\varphi}$
and $\ket{\theta,\varphi}=
  \cos\left( \frac{\theta}{2} \right)\ket{0}+e^{i\varphi}\sin\left(
    \frac{\theta}{2} \right)\ket{1}$. {\bf
    (a)} and {\bf (c)}: Single-qubit rotation around the $x$-axis by angle $\pi$. Here $\ket{0}=\ket{g}_{00}$,
  $\ket{1}=\ket{e}_{00}$, the remaining qubits are initialized in their
  ground state. Only the qubit at $(0,0)$ is coupled to its resonator
  with $E_z=0.8\,{\rm V/\mu m}$ and the drive strength is varied. The ideal gate unitary is $R_x(\pi)=e^{-i\frac{\pi}{2}{\bm\sigma}^x_{00}}$. {\bf (b)} and {\bf (d)}: Two-qubit $\sqrt{i\rm SWAP}$
  gate. Here $\ket{0}=\ket{g}_{00}\ket{e}_{01}$,
  $\ket{1}=\ket{e}_{00}\ket{g}_{01}$, the other qubits are initialized
  in their ground state. The qubits at $(0,0)$ and $(0,1)$ are coupled to
  their resonators with varying but equal field strength $E_z$. The ideal gate
  unitary is $\sqrt{i\rm SWAP}=e^{-i\frac{\pi}{4}(\bm\sigma_{00}^+\bm\sigma_{01}^-+\bm\sigma_{00}^-\bm\sigma_{01}^+)}$. Full curves
  are numerical results obtained by solving the master equation (ME)~(\ref{eq:8}) and dashed curves show analytic
  upper bounds for ideal gates~\cite{supp_mat}.\label{fig:gate_fids}}
\end{figure*}

{\em Nearest neighbor two-qubit gates}. A high fidelity two-qubit gate
can be realized
by a generalization of the resonator-bus mediated qubit-qubit flip-flop
interaction~\cite{Blais-2004a, Blais-2007a, Kloeffel-2013}. Since each qubit is directly coupled only to one
resonator, a virtual photon emitted by the first qubit needs to hop from one
resonator to a neighboring one before being absorbed by the second qubit
(see Fig.~\ref{fig:gate_fids} (b)). A
perturbative analysis~\cite{supp_mat} gives an effective coupling
between qubits at sites $(n, m)$ and $(n', m')$ of
the form
\begin{align}\label{eq:4}
\bm
  H_{XY}&=\sum_{nm,n'm'}K_{nm, n'm'}\bm\sigma_{nm}^+\bm\sigma_{n'm'}^-+{\rm
          h.c.}
\end{align}
where, in the weak coupling regime $J\ll |\omega_Z-\omega_r|$, the coupling strength is given by~\cite{supp_mat}
\begin{align}
  K_{nm,n'm'}&=\frac{(\Delta m+\Delta n)!}{\Delta n!\Delta m!}\frac{\nu_{nm}\nu_{n'm'}}{\Delta}\left( \frac{J}{\Delta}
               \right)^{\Delta m+\Delta n},
\end{align}
with $\Delta n=|n-n'|$, $\Delta m=|m-m'|$ and $\Delta=\omega_r-\omega_Z$. The coupling strength decays
exponentially with distance on the lattice and the nearest
neighbor coupling strength (i.e. for $n'=n$ and $m'=m\pm 1$ or
$n'=n\pm 1$ and $m'=m$), is~\cite{supp_mat}
\begin{align}\label{eq:5}
K_{\rm NN}\simeq
  \frac{J}{\Delta^2}\nu_{nm}\nu_{n'm'}.
\end{align}
Compared with the usual flip-flop interaction strength between two
qubits off-resonantly coupled to the same resonator mode, this coupling
is a factor $J/|\Delta|$ smaller as it involves an additional
off-resonant inter-resonator photon hopping. The interaction~(\ref{eq:4})
acting for a duration $T$, naturally gives rise to the $\sqrt{i\rm SWAP}$ gate when
$K_{\rm NN}T=\pi/4$. Two such gates together with single-qubit
rotations can be used
to implement the CNOT gates required for syndrome measurements in the surface code. As
with the single-qubit gates, it is possible to perform many two-qubit
gates in parallel by taking advantage of the electric field tunability
of the qubit frequency. This is achieved by separating the qubits on the
lattice into two sets with frequencies $\omega_Z^{\rm (r)}$ and
$\omega_Z^{\rm (b)}$ as
illustrated in Fig.~\ref{fig:parallel}. A full syndrome mapping cycle
from the code qubits onto the ancilla qubits can then be performed in four steps.

{\em Readout}. The readout of the ancilla qubits proceeds in
standard fashion by homodyne detection of the dispersive phase
shift incurred by
reflected photons at the bare resonator frequency~\cite{Blais-2007a,
  Koch-2007a}. During readout the code qubits are decoupled from their
resonators. Similar to single-qubit
operations, readout of all ancilla qubits can be performed in
parallel and does not require additional resonators, greatly
simplifying the circuit design. The required reset of the ancilla
qubits to their groundstate after measurement can be implemented for
example by
using the method of~\citet{Geerlings-2013}.

{\em Parameter estimates}. From electrostatic finite element simulations with
optimized cross capacitor designs, we find that
coupling capacitances $C_c$ on the order of a few tens of ${\rm fF}$
can be achieved while strongly suppressing unwanted direct couplings by
more than two orders of magnitude~\cite{supp_mat}. With
a resonator length $l=10\,{\rm mm}$ and capacitance per unit length~\cite{Goppl-2008}
$c=0.127\,{\rm nF/m}$, this leads to a
relative inter-resonator coupling strength $J/(\hbar\omega_r)\simeq
1\%$. For the simulations presented below we take $J/\hbar=159\times 2\pi\,{\rm MHz}$, which
corresponds to a coupling capacitance $C_c\simeq 14\,{\rm fF}$, and set the resonator frequency to $\omega_r=13.35\times 2\pi\,{\rm
  GHz}$. The hole confinement frequency is set to
$\omega_h=28\times 2\pi\,{\rm
  GHz}$, which for an effective hole mass $m\simeq 0.012 m_e$,
where $m_e$ is the electron mass, corresponds to a zero point
fluctuation $x_{\rm ZPF}\simeq 166\,{\rm nm}$. We consider a magnetic field
strength $B_{\perp}=194\,{\rm mT}$, which together
with a zero-field  g-factor for Germanium~\cite{Kloeffel-2013, Watzinger-2016} $g(E_z=0)\simeq 5.5$, corresponds to a
zero-field qubit frequency $\omega_Z(E_z=0)\simeq
14.934\times 2\pi\,{\rm GHz}$. The small length of the nanowire qubits
allows for a coplanar waveguide geometry with a small
trench width, which we take to be $W=0.5\,{\rm \mu m}$. This enhances
the root mean square electric field to about $E_{\rm rms}\simeq
3.73\,{\rm V/m}$. Finally, we assume a Rashba spin-orbit parameter
$\alpha_{\rm DR}/\hbar\simeq 10\, e({\rm nm})^2 \times E_z$. For an applied
field $E_z=1\,{\rm V/\mu m}$, this corresponds to $\ell_{\rm
  SO}\simeq 635\,{\rm nm}$. According to Eq.~(\ref{eq:3}), we thus estimate conservatively that coupling
strengths between $\nu_{nm}=0$ at $E_z=0$ and $\nu_{nm}/\hbar\simeq
40\times 2\pi\,{\rm MHz}$ at $E_z=1\,{\rm V/\mu m}$ are currently
feasible. The corresponding qubit frequency shift
between the ``on'' and the ``off'' states is $\Delta\omega_Z\simeq
1.161\times 2\pi\,{\rm GHz}$, i.e. $\omega_Z(E_z=1\,{\rm V/\mu m})\simeq 13.773\times 2\pi\,{\rm GHz}$, which allows for phase gates on
the nanosecond timescale.

{\em Numerical simulations}. We characterize the theoretical performance of
single and two-qubit gates on a $2\times 2$ lattice in the presence of dissipation and gate
imperfections by
numerically solving the Lindblad master equation (with $\hbar=1$)
\begin{align}\label{eq:8}
\bm{\dot \rho}=-i[\bm H+\bm H_d,\bm \rho]+\kappa\sum_{nm}\mathcal{D}[\bm
  a_{nm}]\bm\rho+\gamma\sum_{nm}\mathcal{D}[\bm\sigma^-_{nm}]\bm\rho.
\end{align}
Here $\bm H$ is given by Eq.~(\ref{eq:1}), $\kappa$ denotes the single
photon loss rate of the resonators, $\gamma=1/T_1$ the qubit decay rate and
$\mathcal{D}[\bm O]\bm\rho=(2\bm O\bm \rho\bm O^{\dagger}-\bm
O^{\dagger}\bm O\bm \rho-\bm\rho\bm O^{\dagger}\bm
O)/2$. Fig.~\ref{fig:gate_fids} (c) shows
the fidelity of a rotation the qubit at lattice site $(0,0)$ around the $x$-axis by angle $\pi$
averaged over all initial states on the Bloch sphere as a function of
the gate duration time $T$ for $\kappa/h=\gamma/h=10\,{\rm kHz}$. This rotation is realized by
a drive on resonator $(0,0)$ of the form $\bm H_d(t)=\varepsilon(t)\left( e^{i\omega_dt}\bm
  a+e^{-i\omega_dt}\bm a^{\dagger} \right)$ with frequency
$\omega_d=\omega_Z+(2\bar n+1)\mychi$ and gaussian envelop
$\varepsilon(t)=\varepsilon\exp[-(t-t_0)^2/(2\sigma^2)]$ with $\varepsilon=\pi\Delta/(2\sigma\nu\sqrt{2\pi})$. Here
$\sigma=T/5$, $t_0=T/2$, $\Delta=\omega_Z-\omega_r$ and $\omega_Z$
is the bare qubit frequency in the ``on'' state with dispersive
shift $\mychi=\nu^2/\Delta$. The
drive frequency shift $(2\bar n+1)\mychi$ with $\bar n\simeq \varepsilon^2/[\Delta^2+(\kappa/2)^2]$
corrects (approximately) for both the Lamb
and Stark shifts. The simulated fidelity (full red curve in
Fig.~\ref{fig:gate_fids} (c)) is upper bounded by $\mathcal{F}_{\varphi}=(1+(1/3)e^{-\gamma T}+(2/3)e^{-\left[
    \gamma/2+\gamma_{\varphi}\right]T})/2$ (dashed curve in
Fig.~\ref{fig:gate_fids} (c)),
which gives the average fidelity for an
ideal gate with a $T_1$-limited qubit subject to photon
shot noise induced dephasing~\cite{Gambetta-2006} with rate
$\gamma_{\varphi}\simeq 2\bar n\kappa(\pi/2)^2$ (blue curve in
Fig.~\ref{fig:gate_fids} (c)). The
difference between the two curves is a measure of gate
imperfections such as deviations from optimal pulse duration and
spurious entanglement between the photons and the qubit, which
increases with the drive strength.

Next we characterize the natural two-qubit gate generated by the
interaction in Eq.~(\ref{eq:4}). Fig.~\ref{fig:gate_fids} (d) shows the
fidelity of a $\sqrt{i{\rm
    SWAP}}=\sqrt{1/2}\,[\openone+\bm{ZZ}-i(\bm{XX}+\bm{YY})]^{1/2}$ gate
between qubits at sites $(0,0)$ and $(0,1)$ obtained for $K_{\rm NN}T=\pi/4$,
averaged over the subset of initial two-qubit states in ${\rm
  span}\,\{\ket{eg},\ket{ge}\}$, while the remaining two qubits are in their ground state. In this case, the gate duration $T$ is
fixed by the interaction strength. The latter however depends on the
strength of the applied electric field $E_z$. For small $E_z$ the
averaged gate fidelity agrees well with that of an ideal $T_1$-limited
$\sqrt{i{\rm SWAP}}$ gate, which for the considered initial states
in the one-excitation manifold, is simply
$\mathcal{F}_0=e^{-\gamma T}$ (dashed curve in
Fig.~\ref{fig:gate_fids} (d)). As the field and
hence the interaction strength is increased, the gate becomes faster
and, at first, the fidelity increases. Because an increasing electric field
also reduces the detuning between the qubit and the resonator, the
dispersive approximation breaks down for too large an applied field, which is
reflected in fluctuations and overall suppression of the fidelity at
strong fields.

{\em Conclusion}. We have proposed a scalable hybrid architecture for
fault tolerant quantum computation via the surface code. The core of
this system consists of a square lattice of capacitively
coupled superconducting resonators, which serves as a two-dimensional
quantum bus to mediate interactions between nanowire hole-spin qubits. By leveraging the
electric tunability of the strong DRSOI, unwanted
couplings between qubits are suppressed. This is a key advantage
compared to other architectures where qubit-qubit
interactions are controlled by frequency tuning and errors due to
spurious off-resonant
couplings scale with the system size. Furthermore, the circuit layout
of our architecture benefits from the small size of the nanowire
qubits and is greatly simplified by efficient component reuse.

{\em Acknowledgments}. This work was supported by the Swiss National
Science Foundation (SNSF) and the NCCR QSIT. D.~L.~acknowledges James
Wootton for useful discussions. The numerical
computations were peformed in a
parallel computing environment at sciCORE (\url{http://scicore.unibas.ch/}) scientific computing core facility
at University of Basel using the Python library QuTip (\url{http://qutip.org/}).
%

\cleardoublepage

\begin{widetext}
\renewcommand{\thefigure}{S\arabic{figure}}  
\begin{center}
{\bf Supplementary Material for\\``Superconducting grid-bus surface
  code architecture for hole-spin qubits''}\vspace{0.2cm}\\
{Simon E. Nigg$^1$, Andreas Fuhrer$^2$ and Daniel
  Loss$^1$}\\
$^1${\small\it Department of Physics, University of Basel,
  Klingelbergstrasse 82, 4056 Basel, Switzerland and}\\
$^2${\small\it IBM Research - Zurich S{\"a}umerstrasse 4, 8803
  R{\"u}schlikon, Switzerland}\\
{\small(Dated: December 21, 2016)}\vspace{0.5cm}\\
\begin{minipage}[ht]{0.8\textwidth}
{\small\hspace{0.3cm}In this supplementary material we provide some details on the
derivation of the effective model describing the grid-bus lattice. We
also discuss the possibility to compactify the architecture via a
procedure that we call ``code folding'' and
present additional results from numerical simulations to supplement
those in the main text.}
\end{minipage}
\end{center}

\section{Derivation of Eqs. (3) and (4)}
\renewcommand{\theequation}{S\arabic{equation}}
Here we derive the effective model, Eq. (4) of the main text. We start
from the Hamiltonian given by the sum of Eqs. (1) and (2) in the main
text which is
\begin{align}
\bm H=\frac{\bm p^2}{2m} +\frac{1}{2}m\omega_h^2\bm x^2 +\alpha_{\rm DR} \bm
  \sigma^y\bm p + \frac{\hbar\omega_Z}{2}\bm\sigma^z+eE_{\rm rms}\bm x(\bm
  a+\bm a^{\dagger})+\hbar\omega_r\bm a^{\dagger}\bm a.
\end{align}
We note that in the absence of a magnetic field, i.e. $\omega_Z=0$,
the spin degree of freedom is conserved as $[\bm \sigma^y,\bm H|_{B=0}]=0$. We
first remove the spin-orbit term by performing a spin conditional
momentum displacement via the unitary operator
\begin{align}
\bm U=\exp\left(  i\frac{\bm x}{\ell_{\rm SO}}\bm\sigma^y
  \right),\quad\ell_{\rm SO}\equiv\frac{\hbar}{m\alpha_{\rm DR}}.
\end{align}
The transformed Hamiltonian is
\begin{subequations}
\begin{align}
\bm{\tilde H}=\bm U\bm H\bm U^{\dagger}&=\frac{\left( \bm
  p-\frac{\hbar}{\ell_{\rm SO}}\bm\sigma^y
  \right)^2}{2m}+\alpha_{\rm DR}\bm\sigma^y\left(\bm p-\frac{\hbar}{\ell_{\rm
                                         SO}}\bm\sigma^y\right)
  +\frac{1}{2}m\omega_h^2\bm x^2 +eE_{\rm rms}\bm x\left( \bm a+\bm
  a^{\dagger} \right)+\hbar\omega_r\bm a^{\dagger}\bm a\\
  &+\frac{\hbar\omega_Z}{2}\left[ \cos\left( \frac{\bm x}{\ell_{\rm SO}}
  \right)+i\sin\left( \frac{\bm x}{\ell_{\rm SO}} \right)\bm\sigma^y
  \right]\bm\sigma^z\left[ \cos\left( \frac{\bm x}{\ell_{\rm SO}}
  \right)-i\sin\left( \frac{\bm x}{\ell_{\rm SO}} \right)\bm\sigma^y
\right]\\
&=\frac{\bm p^2}{2m}+\frac{1}{2}m\omega_h^2\bm x^2+eE_{\rm rms}\bm
  x\left(  \bm a+\bm a^{\dagger} \right)+\hbar\omega_r\bm
  a^{\dagger}\bm a+\frac{\hbar\omega_Z}{2}\left[ \cos\left( \frac{2\bm x}{\ell_{\rm
  SO}} \right)\bm\sigma^z-\sin\left( \frac{2\bm x}{\ell_{\rm SO}}
  \right)\bm\sigma^x\right].
\end{align}
\end{subequations}
In the last equality, we have neglected a c-number term:
$-\hbar^2/(2m\ell_{\rm SO}^2)$. Note that although we have formally
removed the spin-orbit term, the spin and orbital degrees of freedom
of the hole
are now mixed. We are primarily interested in a
situation where the hole occupies the ground state of the harmonic
confinement potential and where the ac field frequency is much lower
than the confinement frequency i.e. $\omega_r\ll\omega_h$, such that
the hole follows the field adiabatically. If further $x_{\rm
  ZPF}=\sqrt{\hbar/(2m\omega_h)}\ll\pi\ell_{\rm SO}$, then the size of
the hole dipole is small
compared with the SOI length and we can expand the trigonometric functions to
leading order yielding
\begin{align}
  \bm{\tilde H}\simeq\frac{\bm p^2}{2m}+\frac{1}{2}m\omega_h^2\bm
  x^2+eE_{\rm rms}\bm
  x\left(  \bm a+\bm a^{\dagger} \right)+\hbar\omega_r\bm
  a^{\dagger}\bm a+\frac{\hbar\omega_Z}{2}\left( \bm\sigma^z-\frac{2\bm x}{\ell_{\rm SO}}
  \bm\sigma^x \right) .
\end{align}
This is Eq.~(3) of the main text. Writing the position and momentum operators of the
hole in second quantized notation as $\bm x =
x_{\rm ZPF}\left( \bm b+\bm b^{\dagger} \right)$
and $\bm p=-\frac{i\hbar}{2x_{\rm ZPF}}\left( \bm b-\bm b^{\dagger}
\right)$, the Hamiltonian becomes (we henceforth drop the $\sim$ and
suppress constant c-number terms)
\begin{align}
\bm H &= \hbar\omega_h\bm b^{\dagger}\bm b+\hbar\omega_r\bm
        a^{\dagger}\bm a+\beta\left( \bm a+\bm
        a^{\dagger} \right)\left( \bm b+\bm b^{\dagger}
        \right)+\frac{\hbar\omega_Z}{2} \bm\sigma^z-\hbar\omega_Z\left( \frac{x_{\rm ZPF}}{\ell_{\rm
        SO}} \right)\left( \bm b+\bm b^{\dagger} \right)\bm\sigma^x,
\end{align}
where we have defined $\beta =eE_{\rm rms}x_{\rm ZPF}$. Further assuming that
$|\omega_Z-\omega_h|\ll|\omega_Z+\omega_h|$, as well as
$|\omega_h-\omega_r|\ll|\omega_r+\omega_h|$, we perform two
rotating wave approximations to neglect counter-rotating terms $\sim\bm b^2,\bm b^{\dagger 2},
\bm a\bm b, \bm a^{\dagger}\bm b^{\dagger}$. Thus
\begin{align}
\bm H &= \hbar\omega_h\bm b^{\dagger}\bm b+\hbar\omega_r\bm
        a^{\dagger}\bm a+\beta\left( \bm a\bm b^{\dagger}+\bm
        a^{\dagger}\bm b \right)+\frac{\hbar\omega_Z}{2} \bm\sigma^z-\hbar\omega_Z\left( \frac{x_{\rm ZPF}}{\ell_{\rm
        SO}} \right)\left( \bm b\bm\sigma^++\bm b^{\dagger}\bm\sigma^- \right),
\end{align}

We next perform the
canonical transformation
\begin{align}
  \bm a &= \cos(\theta)\bm c+\sin(\theta)\bm d,\\
  \bm b &= -\sin(\theta)\bm c+\cos(\theta)\bm d.
\end{align}
The condition that off-diagonal elements in $\bm c$ and $\bm d$ should vanish yields the
equality
\begin{align}
\beta\left(
  \cos^2\theta -\sin^2\theta\right)=\sin\theta\cos\theta\left( \hbar\omega_h-\hbar\omega_r \right)\Leftrightarrow\theta=\frac{1}{2}\arctan\left[ \frac{2\beta}{\hbar\omega_h-\hbar\omega_r} \right].
\end{align}
In this basis, the Hamiltonian reads
\begin{align}
\frac{\bm H}{\hbar}=\omega_d\bm d^{\dagger}\bm d+\omega_c\bm
  c^{\dagger}\bm c+\frac{\omega_Z}{2}\bm \sigma^z+\omega_Z\sin\left( \theta \right)\left( \frac{
  x_{\rm ZPF}}{\ell_{\rm SO}} \right)\left(
  \bm c^{\dagger}\bm\sigma^-+\bm c\bm\sigma^+ \right)-\omega_Z\cos(\theta)\left(
  \frac{x_{\rm ZPF}}{\ell_{\rm SO}} \right)\left(\bm
  d^{\dagger}\bm\sigma^-+\bm d\bm\sigma^+  \right).
\end{align}
Here
\begin{align}
\hbar\omega_d&=\hbar\omega_r\sin^2\theta+\hbar\omega_h\cos^2\theta+2\beta\sin\theta\cos\theta,\\
\hbar\omega_c&=\hbar\omega_r\cos^2\theta+\hbar\omega_h\sin^2\theta-2\beta\sin\theta\cos\theta.
\end{align}
We assume that $\beta\ll\hbar|\omega_h-\omega_r|$ and consequently approximate $\theta\simeq
\beta/(\hbar\omega_h-\hbar\omega_r)$ so that $\cos(\theta)\simeq 1$
and $\sin(\theta)\simeq \beta/(\hbar\omega_h-\hbar\omega_r)$ and
$\hbar\omega_d\simeq\hbar\omega_h+2\beta^2/(\hbar\omega_h-\hbar\omega_r)\simeq\hbar\omega_h$
as well as
$\hbar\omega_c\simeq\hbar\omega_r-2\beta^2/(\hbar\omega_h-\hbar\omega_r)\simeq\hbar\omega_r$. The
Hamiltonian is then
\begin{align}
\frac{\bm H}{\hbar}\simeq\omega_h\bm d^{\dagger}\bm d+\omega_r\bm
  c^{\dagger}\bm c+\frac{\omega_Z}{2}\bm \sigma^z+\frac{\beta \omega_Z}{\hbar(\omega_h-\omega_r)}\left( \frac{
  x_{\rm ZPF}}{\ell_{\rm SO}} \right)\left(
  \bm c^{\dagger}\bm\sigma^-+\bm c\bm\sigma^+ \right)-\omega_Z\left(
  \frac{x_{\rm ZPF}}{\ell_{\rm SO}} \right)\left(\bm
  d^{\dagger}\bm\sigma^-+\bm d\bm\sigma^+  \right).
\end{align}

Next, we focus on the regime where
\begin{align}
\frac{\omega_h}{\omega_Z}-1\gg\frac{x_{\rm ZPF}}{\ell_{\rm SO}}.
\end{align}

and perform a Schieffer-Wolff transformation with
\begin{align}
\bm U_{\rm
  SW}=\exp\left[\frac{\omega_Z}{\omega_h-\omega_Z}\left( \frac{x_{\rm
      ZPF}}{\ell_{\rm SO}} \right)\left( \bm
  d^{\dagger}\bm\sigma^--\bm d\bm\sigma^+ \right)\right].
\end{align}

To leading order we find
\begin{align}
\bm H\simeq \hbar\omega_h\bm d^{\dagger}\bm d+\mychi\bm d^{\dagger}\bm
  d\bm\sigma^z+\hbar\omega_r\bm c^{\dagger}\bm c+\frac{\hbar\omega_Z^\prime}{2}\bm\sigma^z+\frac{\beta
  \omega_Z}{\omega_h-\omega_r}\left( \frac{x_{\rm ZPF}}{\ell_{\rm
  SO}} \right)\left( \bm c^{\dagger}\bm\sigma^-+\bm c\bm\sigma^+ \right),
\end{align}
where
\begin{align}
\hbar\omega_Z^\prime&\simeq \hbar\omega_Z-\mychi,\,\quad\mychi\simeq\frac{\hbar\omega_Z^2}{\omega_h-\omega_Z}\left( \frac{x_{\rm
        ZPF}}{\ell_{\rm SO}} \right)^2.
\end{align}
As stated previously, we are interested in the situation where the
hole remains in the ground state of the confinement potential. At zero
temperature we can then neglect the dynamics of the $\bm d$ mode and obtain the final
effective model for the hole-spin qubit coupled to the resonator mode
(note that $\bm c\simeq\bm a$ in the considered regime)
\begin{align}\label{eq:S16}
\bm H_{\rm eff} = \hbar\omega_r\bm a^{\dagger}\bm a+\frac{\hbar\omega_Z^\prime}{2}\bm\sigma^z+\nu\left( \bm a^{\dagger}\bm\sigma^-+\bm a\bm\sigma^+ \right),
\end{align}
with coupling strength
\begin{align}
\nu=\frac{\beta
  \omega_Z}{\omega_h-\omega_r}\left( \frac{x_{\rm ZPF}}{\ell_{\rm
  SO}} \right),
\end{align}
and renormalized qubit frequency
\begin{align}
\omega_Z'=\omega_Z\left[ 1-\frac{\omega_Z}{\omega_h-\omega_Z}\left(
  \frac{x_{\rm ZPF}}{\ell_{\rm SO}} \right)^2 \right].
\end{align}
These are Eqs.~(4) to (6) of the main text.
A few comments are in order. Notice first that via the ``Lamb'' shift
the effective Zeeman energy $\hbar\omega_Z^\prime$, which determines the qubit
frequency depends quadratically on the external electric field. As
observed in the main text, this means that the ``off'' state, $E_z=0$,
is a sweet-spot where the qubit is protected against small electric
field fluctuations to linear order. Second, note that both the qubit frequency and the
qubit-field coupling are proportional to the applied magnetic
field. These results are consistent with those of~\citet{Kloeffel-2013}.

\section{Coupled microwave resonator lattice}
\begin{figure}[ht]
\includegraphics[width=0.3\textwidth]{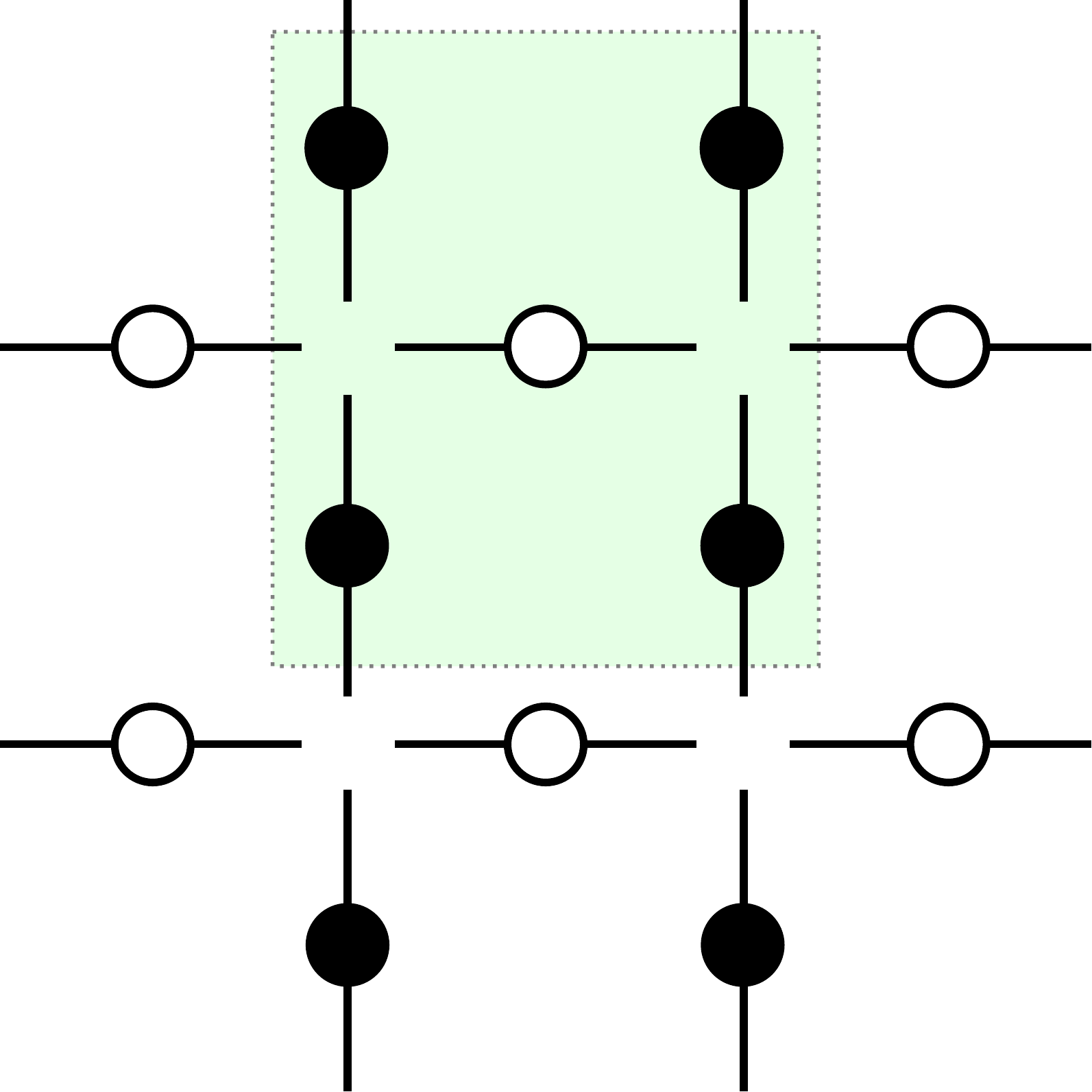}\hfill\includegraphics[width=0.7\textwidth]{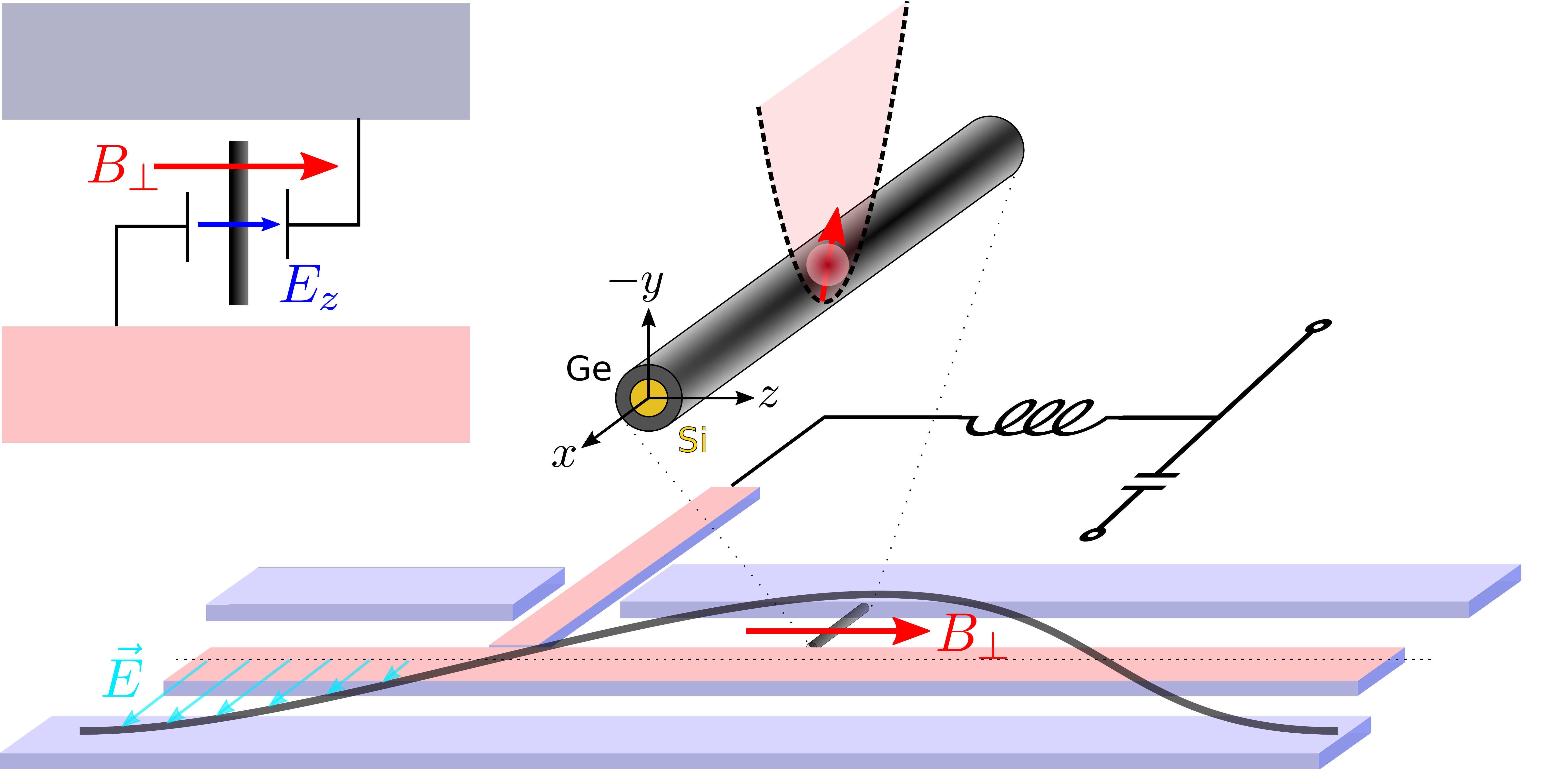}\caption{{\bf
  Left
  panel:} schematics of the resonator-qubit lattice. Black and white dots
  represent qubits while straight lines represent microwave resonators. {\bf Right
  panel:} Schematics
  of a CPW resonator with voltage biased center pin. The hole-spin qubit
  lives in a quantum dot formed in the wire either during growth or
  induced by additional gates (not shown). An electric
  field perpendicular to the wire is controlled by biasing the
  center conductor of the resonator. Furthermore, a magnetic field is
  applied in the plane of the resonators at an angle to the nanowires so as to generate
  a component perpendicular to all the nanowires. \label{fig:CPW}}
\end{figure}
Here we consider how to couple individual qubits together as required
by the surface code. The resonator lattice system is depicted schematically in Fig.~\ref{fig:CPW}
(left panel). Note that because of the strong suppression of the $g$-factor
along the axis of the nanowires~\cite{Maier-2013}, we can neglect the
component of the magnetic field along the nanowires. The magnetic
field can be applied either perpendicular to the plane of the
resonators or (preferably) in-plane. Here we focus on the latter
situation. Hence each
isolated site
of the lattice is described by (\ref{eq:S16}).

We first focus on the Hamiltonian for the coupled resonators without the
qubits. The novel feature here is the four-way capacitive coupling. A
capacitor design which maximizes the capacitance between resonators at
a right angle to each other and at the same times minimizes the direct
capacitance across the junction is shown in Fig.~\ref{fig:capa_design} alongside with results from numerical finite
element simulations, where we plot the different
capacitances of the structure as a function of the capacitor lengths
for varying channel widths. As this example illustrates, coupling
capacitances on the order of a few tens of femto Farrads are feasible
while achieving strong suppression of unwanted capacitances by more
than two orders of magnitude.
\begin{figure}[ht]
\includegraphics[width=0.45\textwidth]{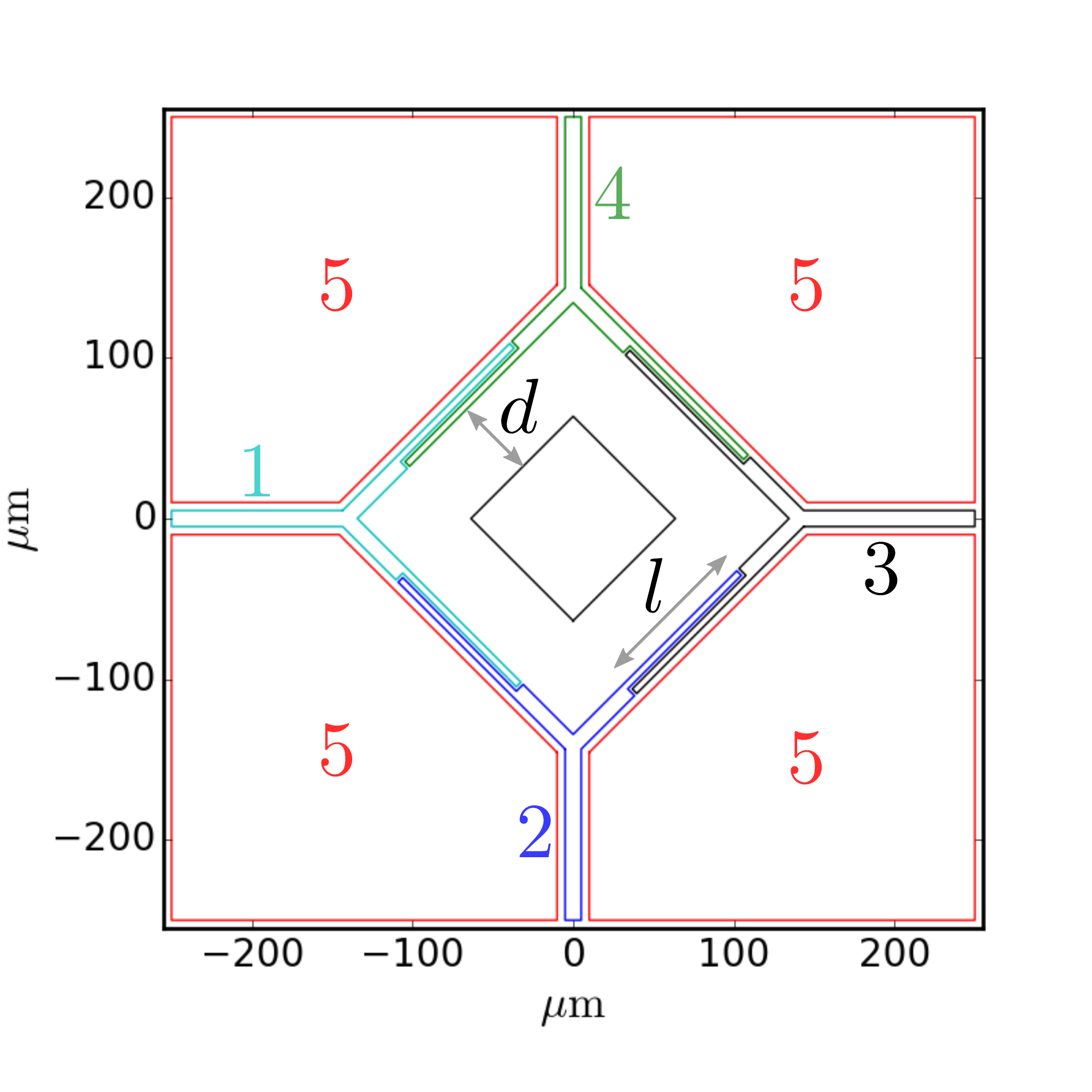}\hfill\includegraphics[width=0.55\textwidth]{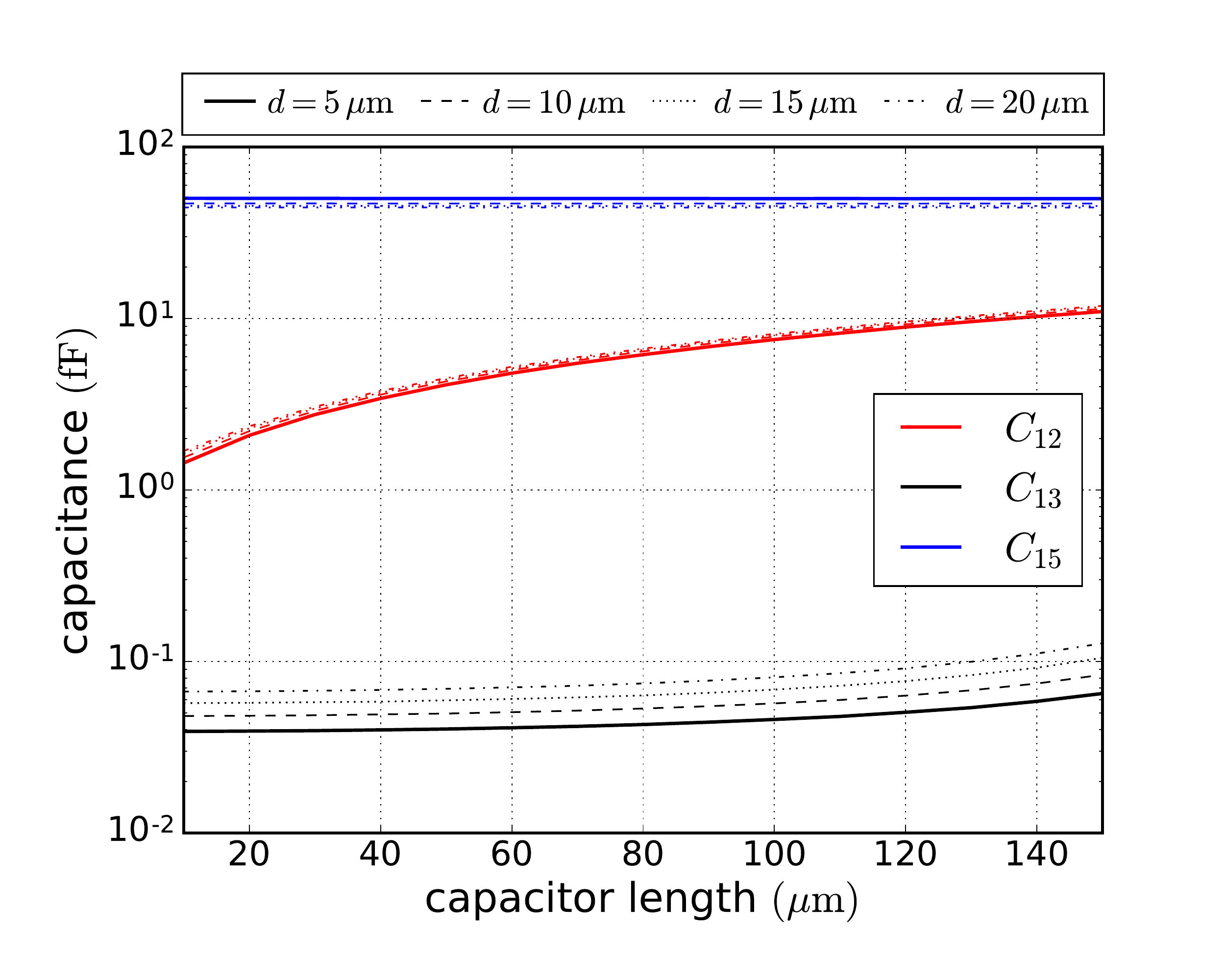}\caption{Capacitor
design to maximize capacitive coupling between resonators at a right
angle to each other while minimizing capacitive coupling directly
across. {\bf Left panel}: Four-way capacitor structure. {\bf right
  panel}: Finite element simulation results.\label{fig:capa_design}}
\end{figure}

Motivated by this we neglect the direct cross capacitance and model 
the coupling of one resonator to its four perpendicular neighbors (see
Fig.~1 (g) of the main text) with
the Lagrangian
\begin{align}\label{eq:S2}
\mathcal{L}=\frac{1}{2}\vec{\dot\varphi}^T\bm C\vec{\dot\varphi}-\frac{1}{2}\sum_{n=1}^5\frac{\varphi_n^2}{L}.
\end{align}
Here we have modeled each isolated resonator mode as a single $LC$ resonance
and $\varphi_n$ denotes the phase variable associated with the $n$-th
resonator in units of the reduced flux quantum $\hbar/(2e)$. We have
further introduced the real symmetric $5\times 5$ capacitance
matrix $\bm C$. For a symmetric arrangement of equal cavities and denoting
with $C$ the self capacitance of each center conductor and $C_c$ the
pairwise coupling capacitances, the capacitance matrix reads
\begin{align}
\bm C = \begin{pmatrix}
C+4C_c&-C_c&-C_c&-C_c&-C_c\\
-C_c&C+4C_c&-C_c&-C_c&-C_c\\
-C_c&-C_c&C+4C_c&-C_c&-C_c\\
-C_c&-C_c&-C_c&C+4C_c&-C_c\\
-C_c&-C_c&-C_c&-C_c&C+4C_c
\end{pmatrix}.
\end{align}
The associated quantum
Hamiltonian, obtained by canonical quantization is given by
\begin{align}
\bm H_5 = \omega_r\bm a_n^{\dagger}\bm
  a_n^{}+J\left( \bm a_n^{\dagger}\bm a_{n-1}^{}+ \bm a_n^{\dagger}\bm a_{n+1}^{}+ \bm a_{n+1}^{\dagger}\bm a_{n}^{}+ \bm a_{n-1}^{\dagger}\bm a_{n}^{} \right),
\end{align}
with $\omega_r=1/\sqrt{L(C+4C_c)}$ and $J=2\hbar\omega_r
\frac{C_c}{C+4C_c}$.

The Hamiltonian for an entire
square lattice of resonators with nearest neighbor capacitive coupling is then given
by a straightforward generalization
\begin{align}
\bm H &=\omega_r\sum_{n,m=1}^N\bm
  a_{nm}^{\dagger}\bm a_{nm}^{}+J\sum_{n,
  m}\left( \bm a_{nm}^{\dagger}\bm a_{nm+1}^{}+ \bm
  a_{nm}^{\dagger}\bm a_{nm-1}^{}+ \bm a_{nm}^{\dagger}\bm
  a_{n+1m}^{}+ \bm a_{nm}^{\dagger}\bm a_{n-1m}^{}+{\rm h.c.}\right).
\end{align}

Including the spin qubits and their coupling to the resonators
immediately leads to the Jaynes-Cummings-Hubbard model (JCH) of Eq.~(7) in
the main text.

\section{Qubit gates}
Here we provide further details on the single and two-qubit gates.
\subsection{$R_z$ gate}
A rotation of a selected qubit at lattice point $(n,m)$ around the $z$ axis can be achieved by
exploiting the electric field dependence of the qubit frequency:
If the perpendicular
electric field in resonator at $(n,m)$ is tuned via the center pin to the
working point $E_z^*$, then the qubit resonance frequency is
decreased by an amount $\Delta\omega_{nm}=\omega_Z(E_z=0)-\omega_Z(E_z=E_z^*)$. By simply waiting for a time $T$, a rotation
$R_z(T\Delta\omega_{nm})$ around $z$ is induced. For the parameters
given in the main text and a ``on'' field strength of $E_z^*=1\,{\rm
  V/\mu m}$ for example, $\Delta\omega_{nm}=1.161\times 2\pi\,{\rm GHz}$ which
translates to a $\pi$ rotation around $z$ in a time $T_{\pi}\simeq
430\,{\rm ps}$.

\subsection{Analytic fidelity upper bounds}
In this section we derive the analytic upper bounds for the average fidelities of
single and two-qubit gates given in the main text and shown in Fig.~3
(c) and (d) of the main text. We start with single qubit rotations. If the gate were perfect and the only source of
imperfection were due to decoherence, then at zero temperature the single qubit density
matrix would evolve as
\begin{align}
\bm\rho(0)\rightarrow\bm \rho(T)=\begin{pmatrix}\rho_{ee}e^{-\gamma
    T}&\rho_{eg}e^{-\left(
      \frac{\gamma}{2}+\gamma_{\varphi}\right)T}\\\rho_{ge}e^{-\left(
      \frac{\gamma}{2}+\gamma_{\varphi}\right)T}&1+(\rho_{gg}-1)e^{-\gamma
  T}\end{pmatrix},
\end{align}
where $\rho_{ij}$ are the components of the initial state,
$\gamma$ is the relaxation rate and $\gamma_{\varphi}$ the
dephasing rate. The
fidelity for a given initial state $\ket{\theta,\varphi}=\cos\left(
  \frac{\theta}{2} \right)\ket{0}+e^{i\varphi}\sin\left(
  \frac{\theta}{2} \right)\ket{1}$ on the Bloch sphere is given by
\begin{align}
\mathcal{F}(\theta,
  \varphi)=\braket{\theta,\varphi|\bm\rho(t)|\theta,\varphi}=\left(
  \sin^4(\theta/2)+\cos^4(\theta/2) \right)e^{-\gamma
  T}+2\sin^2(\theta/2)\cos^2(\theta/2)e^{-[\frac{\gamma}{2}+\gamma_{\varphi}]T}+\cos^2(\theta/2)(1-e^{-\gamma T}).
\end{align}
Averaging over the Bloch sphere then yields
\begin{align}
\mathcal{F}_{\varphi}=\frac{1}{4\pi}\int_0^{2\pi}d\varphi\int_0^{\pi}d\theta\sin(\theta)\mathcal{F}(\theta,\varphi)=\frac{1}{2}\left(
  1+\frac{1}{3}e^{-\gamma T}+\frac{2}{3}e^{-\left[ \frac{\gamma}{2}+\gamma_{\varphi} \right]T} \right).
\end{align}
Note that in our situation $T$ is the gate time, which is inversely
proportional to the drive strength. Hence, $\gamma_{\varphi}$, which in our
case is determined by photon shot noise, itself depends on $T$
(see Fig.~3 (c) of the main text).

For the two-qubit flip-flop gate we focus on a subset of initial states given by
${\rm span}\left\{ \ket{eg}, \ket{ge} \right\}$. The reason is that
for a perfect gate, the flip-flop interaction acts trivially on both
$\ket{gg}$ and $\ket{ee}$. The two states $\ket{eg}$ and $\ket{ge}$
span an effective Bloch sphere. Because these states live in the one
excitation manifold, the fidelity of an ideal gate does in
this case not depend on the angles $\theta$ and $\varphi$ and is thus
simply given by $\mathcal{F}_0=e^{-\gamma T}$. Note that in this case
we do not have photon induced dephasing to leading order since the resonator remains in
the vacuum state.
\section{Flip-flop interactions on the grid-bus lattice}
Here we derive the effective flip-flop interaction between neighboring
qubits on a $N\times M$
lattice of coupled resonators. It is convenient to
introduce the Fourier modes
\begin{align}
\bm a_{nm}^{}=\frac{1}{\sqrt{NM}}\sum_{k=0}^{N-1}\sum_{l=0}^{M-1}e^{i\frac{2\pi
  k}{N}n}e^{i\frac{2\pi l}{N}m}\bm b_{kl}^{},
\end{align}
with $[\bm b_{kl}^{},\bm b_{k'l'}^{\dagger}]=\delta_{kk'}\delta_{ll'}$, all other commutators
being zero. In this basis, the JCH Hamiltonian reads
\begin{subequations}
\begin{align}
\bm H &=
  \sum_{n,m}\frac{\omega_{nm}}{2}\bm\sigma_{nm}^z+\sum_{k,l}\left(
        \omega_r+2J\left[ \cos\left( \frac{2\pi k}{N} \right)
        +\cos\left( \frac{2\pi l}{M} \right)\right]\right)\bm
  b_{kl}^{\dagger}\bm b_{kl}^{}\label{eq:S3}\\
&+\frac{1}{\sqrt{NM}}\sum_{k,l}\sum_{n,m}\nu_{nm}\left( e^{i\frac{2\pi
  k}{N}n}e^{i\frac{2\pi l}{N}m}\bm
  b_{kl}^{}\bm\sigma_{nm}^++e^{-i\frac{2\pi
  k}{N}n}e^{-i\frac{2\pi l}{N}m}\bm b_{kl}^{\dagger}\bm \sigma_{nm}^-
  \right).\label{eq:S4}
\end{align}
\end{subequations}
We next consider the dispersive regime. We define the anti-hermitian operator
\begin{align}
\bm
  S=\frac{1}{\sqrt{NM}}\sum_{nm}\sum_{kl}\frac{\nu_{nm}}{\omega_{nm}-\omega_r-2J\left[
  \cos\left( \frac{2\pi k}{N} \right) +\cos\left( \frac{2\pi l}{M}
  \right)\right]}\left(e^{-i\frac{2\pi
  k}{N}n}e^{-i\frac{2\pi l}{M}m} \bm\sigma_{nm}^-\bm
  b_{kl}^{\dagger}-e^{i\frac{2\pi
  k}{N}n}e^{i\frac{2\pi l}{M}m}\bm\sigma_{nm}^{+}\bm b_{kl} \right).
\end{align}
Splitting the Hamiltonian as $\bm H=\bm H_0+\bm V$, with $\bm H_0$
given by (\ref{eq:S3}) and $V$ given by~(\ref{eq:S4}), we have
\begin{align}
[\bm H_0, \bm S]=-\bm V.
\end{align}
In the dispersive regime of interest  we have
$|\omega_{nm}-\omega_{kl}|\gg \nu_{nm}$, where
$\omega_{kl}=\omega_r+2J\left[ \cos\left( \frac{2\pi k}{N} \right)
  +\cos\left( \frac{2\pi l}{M} \right)\right]$.
The leading order correction to $\bm H_0$ upon performing
the Schrieffer-Wolff transformation $\bm H\rightarrow e^{-\bm S}\bm
He^{\bm S}$, is given by $(1/2)[\bm V, \bm S]$ where
\begin{subequations}
\begin{align}
[\bm V,\bm S]&=\frac{1}{NM}\sum_{nm,n'm'}\sum_{kl,k'l'}\left\{ 
\frac{\nu_{nm}\nu_{n'm'}e^{i\frac{2\pi k}{N}n}e^{i\frac{2\pi
               l}{M}m}e^{-i\frac{2\pi k'}{N}n'}e^{-i\frac{2\pi
               l'}{M}m'}}{\omega_{n'm'}-\omega_{k'l'}}\left[ \bm
               b_{kl}^{}\bm\sigma_{nm}^{+},\bm\sigma_{n'm'}^{-}\bm
               b_{k'l'}^{\dagger} \right] \right\}\label{eq:S7},\\
&-\frac{1}{NM}\sum_{nm,n'm'}\sum_{kl,k'l'}\left\{ 
\frac{\nu_{nm}\nu_{n'm'}e^{-i\frac{2\pi k}{N}n}e^{-i\frac{2\pi
               l}{M}m}e^{i\frac{2\pi k'}{N}n'}e^{i\frac{2\pi
               l'}{M}m'}}{\omega_{n'm'}-\omega_{k'l'}}\left[ \bm
               b_{kl}^{\dagger}\bm\sigma_{nm}^{-},\bm\sigma_{n'm'}^{+}\bm
               b_{k'l'}^{} \right] \right\}\label{eq:S8}.
\end{align}
\end{subequations}
We further have
\begin{subequations}
\begin{align}
[\bm b_{kl}^{}\bm\sigma_{nm}^+,\bm\sigma_{n'm'}^-\bm
  b_{k'l'}^{\dagger}]&=\delta_{nn'}\delta_{mm'}\bm b_{kl}^{}\bm
                       b_{k'l'}^{\dagger}\bm\sigma_{nm}^z+\delta_{kk'}\delta_{ll'}\bm\sigma_{nm}^{+}\bm\sigma_{n'm'}^{-}\label{eq:S5},\\
[\bm b_{kl}^{\dagger}\bm\sigma_{nm}^-,\bm\sigma_{n'm'}^+\bm
  b_{k'l'}^{}]&=-\delta_{nn'}\delta_{mm'}\bm b_{kl}^{\dagger}\bm
                       b_{k'l'}^{}\bm\sigma_{nm}^z-\delta_{kk'}\delta_{ll'}\bm\sigma_{nm}^{-}\bm\sigma_{n'm'}^{+}\label{eq:S6}.
\end{align}
\end{subequations}
Substituting, we find after some algebra that the effective
Hamiltonian can be written as
\begin{align}
\bm H = \bm H_0+\bm H_{\rm L}+\bm H_{\rm d} +\bm H_{\rm nd}+\bm H_{XY},
\end{align}
with
\begin{subequations}
\begin{align}
\bm H_{\rm L}&=\frac{1}{2NM}\sum_{nm}\sum_{kl}\frac{\nu_{nm}^2}{\omega_{nm}-\omega_{kl}}\bm\sigma_{nm}^z,\label{eq:S11}\\
\bm H_{\rm
  d}&=\frac{1}{NM}\sum_{nm}\sum_{kl}\frac{\nu_{nm}^2}{\omega_{nm}-\omega_{kl}}\bm\sigma_{nm}^z\bm
  b_{kl}^{\dagger}\bm b_{kl}^{},\label{eq:S9}\\
\bm H_{\rm
  nd}&=\frac{1}{2NM}\sum_{nm}\sum_{k\not=k',l\not=l'}\frac{\nu_{nm}^2}{\omega_{nm}-\omega_{k'l'}}e^{i\frac{2\pi}{N}(k-k')n}e^{i\frac{2\pi}{M}(l-l')m}\bm\sigma_{nm}^z\bm
       b_{kl}^{}\bm b_{k'l'}^{\dagger}+{\rm h.c.},\label{eq:S10}\\
\bm
  H_{XY}&=\frac{1}{NM}\sum_{nm,n'm'}\sum_{kl}\frac{\nu_{nm}\nu_{n'm'}}{\omega_{n'm'}-\omega_{kl}}e^{i\frac{2\pi}{N}k(n-n')}e^{i\frac{2\pi}{M}l(m-m')}\bm\sigma_{nm}^+\bm\sigma_{n'm'}^-+{\rm
          h.c.}\label{eq:S14}
\end{align}
\end{subequations}
The first term~(\ref{eq:S11}) corresponds to a Lamb shift
renormalization of the qubit frequencies. In the regime where
$|\omega_{nm}-\omega_r|=|\Delta_{nm}|>4J$, the qubit frequency shift
can be upper bounded as follows
\begin{align}
|\Delta\omega_{nm}|&=\left|\frac{\nu_{nm}^2}{NM\Delta_{nm}}\sum_{kl}\frac{1}{1-\frac{2J}{\Delta_{nm}}\left[
  \cos\left( \frac{2\pi k}{N} \right) +\cos\left( \frac{2\pi l}{M}
  \right)\right]}\right|\\
&=\left|\frac{\nu_{nm}^2}{NM\Delta_{nm}}\sum_{kl}\sum_{s=0}^{\infty}\left(
  \frac{2J}{\Delta_{nm}} \right)^s\left[\cos\left( \frac{2\pi}{N}k
  \right)+\cos\left( \frac{2\pi}{M}l \right)  \right]^s\right|\\
&\leq\frac{\nu_{nm}^2}{|\Delta_{nm}|-4J }.
\end{align}
The term $\bm H_{\rm d}$
corresponds to the dispersive interaction between the qubit and
the eigenmodes of the coupled resonators. The term $\bm H_{\rm nd}$
corresponds to a qubit-state-dependent inter-eigenmode hopping term
and finally, $\bm H_{XY}$ corresponds to a virtual photon mediated
flip-flop interaction between qubit pairs.

\subsection{Coupling range in the weak coupling limit}
We next elucidate the form of the two-qubit interaction and its
range. For simplicity we consider the case of equal sub-systems such
that $\nu_{nm}=\nu$, $\omega_{nm}=\omega_Z$. We further define
$\Delta=\omega_Z-\omega_r$. Then the matrix elements of (\ref{eq:S14}) are
\begin{align}
  K_{nm,n'm'}=\frac{1}{NM}\frac{\nu^2}{\Delta}\sum_{k,l}\frac{1}{1-\frac{2J}{\Delta}\left[
  \cos\left( \frac{2\pi k}{N} \right) +\cos\left( \frac{2\pi l}{M} \right)\right]}e^{i\frac{2\pi}{N}k(n-n')}e^{i\frac{2\pi}{M}l(m-m')}.
\end{align}
We consider the weak coupling regime where $2J<|\Delta|$. Then
resolving the geometric series we have
\begin{align}
  K_{nm,n'm'}=\frac{1}{NM}\frac{\nu^2}{\Delta}\sum_{k,l}\sum_{s=0}^{\infty}\left( \frac{2J}{\Delta} \right)^s\left[
  \cos\left( \frac{2\pi k}{N} \right) +\cos\left( \frac{2\pi l}{M} \right)\right]^se^{i\frac{2\pi}{N}k(n-n')}e^{i\frac{2\pi}{M}l(m-m')}.
\end{align}
Further using the binomial formula we obtain
\begin{align}
  K_{nm,n'm'}=\frac{1}{NM}\frac{\nu^2}{\Delta}\sum_{s=0}^{\infty}\sum_{q=0}^s\sum_{k,l}{{s}\choose{q}}\left( \frac{2J}{\Delta} \right)^s  \cos^q\left( \frac{2\pi k}{N} \right)\cos^{s-q}\left( \frac{2\pi l}{M} \right)e^{i\frac{2\pi}{N}k(n-n')}e^{i\frac{2\pi}{M}l(m-m')}.
\end{align}
Further writing $\cos(x)=(e^{ix}+e^{-ix})/2$ and applying the binomial
formula twice more we find
\begin{align}
  K_{nm,n'm'}&=\frac{1}{NM}\frac{\nu^2}{\Delta}\sum_{s=0}^{\infty}\sum_{q=0}^s\sum_{r=0}^q\sum_{t=0}^{s-q}\sum_{k,l}{{s}\choose{q}}{{q}\choose{r}}{{s-q}\choose{t}}\left(
  \frac{2J}{\Delta} \right)^s  \frac{1}{2^q}\frac{1}{2^{s-q}}e^{\frac{2\pi i
  kr}{N}}e^{-\frac{2\pi i k(q-r)}{N}}e^{\frac{2\pi i l
  t}{M}}e^{-\frac{2\pi i
  l(s-q-t)}{M}}e^{i\frac{2\pi}{N}k(n-n')}e^{i\frac{2\pi}{M}l(m-m')}\\
&=\frac{1}{NM}\frac{\nu^2}{\Delta}\sum_{s=0}^{\infty}\sum_{q=0}^s\sum_{r=0}^q\sum_{t=0}^{s-q}{{s}\choose{q}}{{q}\choose{r}}{{s-q}\choose{t}}\left(
  \frac{J}{\Delta} \right)^s \sum_{k,l} e^{\frac{2\pi
  i}{N}(2r-q+n-n')k}e^{\frac{2\pi i}{M}(2t-s+q+m-m')l}\\
&=\frac{\nu^2}{\Delta}\sum_{s=0}^{\infty}\sum_{q=0}^s\sum_{r=0}^q\sum_{t=0}^{s-q}\frac{s!}{r!(q-r)!t!(s-q-t)!}\left(
  \frac{J}{\Delta} \right)^s \delta_{2r-q+n-n',0}\,\delta_{2t-s+q+m-m',0}.
\end{align}
To make further progress, we consider the different cases. Without
restriction of generality we let $\Delta n=n-n'\geq 0$ as well as
$\Delta m=m-m'\geq 0$. There are four possible cases:
\begin{itemize}
\item[(i)] $\Delta n$ and $\Delta m$ even. Then the Kronecker deltas
  imply that only terms with even $q$ and even $s$ will
  contribute. Hence we set $s=2k$ and $q=2l$. We have
\begin{align}
  K_{nm,n'm'}&=\frac{\nu^2}{\Delta}\sum_{k=(\Delta m+\Delta n)/2}^{\infty}\sum_{l=\Delta
               n/2}^{k-\Delta m/2}\frac{(2k)!}{\left(l-\frac{\Delta
               n}{2}\right)!\left(l+\frac{\Delta
               n}{2}\right)!\left(k-l-\frac{\Delta m}{2}\right)!\left(k-l+\frac{\Delta m}{2}\right)!}\left( \frac{J}{\Delta} \right)^{2k}.
\end{align}

\item[(ii)] $\Delta n$ and $\Delta m$ odd. Then $q$ must be
  odd while $s$ must be even. Writing $s=2k$ and $q=2l+1$, we have
\begin{align}
  K_{nm,n'm'}&=\frac{\nu^2}{\Delta}\sum_{k=(\Delta m+\Delta n)/2}^{\infty}\sum_{l=(\Delta
               n-1)/2}^{k-(1+\Delta m)/2}\frac{(2k)!}{\left(l+\frac{1-\Delta
               n}{2}\right)!\left(l+\frac{1+\Delta
               n}{2}\right)!\left(k-l-\frac{1+\Delta m}{2}\right)!\left(k-l-\frac{1-\Delta m}{2}\right)!}\left( \frac{J}{\Delta} \right)^{2k}.
\end{align}

\item[(iii)] $\Delta n$ even and $\Delta m$ odd. Then $q$ must be even
  and $s$ must be odd. Writing $s=2k+1$ and $q=2l$ we have
\begin{align}
  K_{nm,n'm'}&=\frac{\nu^2}{\Delta}\sum_{k=(\Delta m+\Delta n-1)/2}^{\infty}\sum_{l=\Delta
               n/2}^{k+(1-\Delta m)/2}\frac{(2k+1)!}{\left(l-\frac{\Delta
               n}{2}\right)!\left(l+\frac{\Delta
               n}{2}\right)!\left(k-l+\frac{1-\Delta m}{2}\right)!\left(k-l+\frac{1+\Delta m}{2}\right)!}\left( \frac{J}{\Delta} \right)^{2k+1}.
\end{align}

\item[(iv)] $\Delta n$ odd and $\Delta m$ even. Then both $q$ and $s$ must be odd. Writing $q=2l+1$ and $s=2k+1$ we have
\begin{align}
  K_{nm,n'm'}&=\frac{\nu^2}{\Delta}\sum_{k=(\Delta m+\Delta n-1)/2}^{\infty}\sum_{l=(\Delta
               n-1)/2}^{k-\Delta m/2}\frac{(2k+1)!}{\left(l+\frac{1-\Delta
               n}{2}\right)!\left(l+\frac{1+\Delta
               n}{2}\right)!\left(k-l+\frac{1-\Delta m}{2}\right)!\left(k-l+\frac{1+\Delta m}{2}\right)!}\left( \frac{J}{\Delta} \right)^{2k+1}.
\end{align}
\end{itemize}

In all four cases, when $J\ll\Delta$, the leading order term is
\begin{align}
  K_{nm,n'm'}&=\frac{(\Delta m+\Delta n)!}{\Delta n!\Delta m!}\frac{\nu^2}{\Delta}\left( \frac{J}{\Delta}
               \right)^{\Delta m+\Delta n}.
\end{align}


Hence to leading order, the coupling strength decays exponentially
with the distance between the involved resonators as measured by
$\Delta n+\Delta m=|n-n'|+|m-m'|$. By appropriately biasing the center conductors of adequate
pairs of neighboring cavities, we can implement the pairwise two-qubit
interactions between code and ancilla qubits required for the surface
code. Note that, importantly, if we choose two different sets of
frequencies for the white and black qubits illustrated in Fig.~\ref{fig:CPW}, we can realize these operations in
parallel as explained in the main text (see also Fig. 2 in the main text).

To leading order, the qubit-qubit interaction is of the standard $XX+YY$ type:
\begin{align}
\bm H_{XY}=\sum_{nn',mm'}K_{nn',mm'}\left( \bm\sigma_{nm}^+\bm\sigma_{n'm'}^-+ \bm\sigma_{nm}^+\bm\sigma_{n'm'}^-\right)=\frac{1}{2}\sum_{nn',mm'}K_{nn',mm'}\left( \bm\sigma_{nm}^x\bm\sigma_{n'm'}^x+ \bm\sigma_{nm}^y\bm\sigma_{n'm'}^y\right)
\end{align}

For two qubits with equal couplings and frequencies, coupled to nearest neighbor
lattices (either $m=m'\pm 1$, $n=n'$ or $n=n'\pm 1$, $m=m'$) the
coupling strength is
\begin{align}
  K_{\rm NN}\simeq J\left( \frac{\nu}{\Delta} \right)^2.
\end{align}
The XY interaction can be used to implement the $i{\rm SWAP}$
gate. Indeed, since $[\bm X\bm X,\bm Y\bm Y]=0$ and $(\bm X\bm
X)^2=(\bm Y\bm Y)^2=\openone_2\otimes\openone_2$, one has (setting $A=J_{\rm
    NN}/2$)
\begin{align}
e^{-iAt(\bm X\bm X+\bm Y\bm Y)}&=e^{-iAt\bm X\bm X}e^{-iAt\bm Y\bm Y}=\left(\cos\left(
  At \right)-i\sin\left( At \right)\bm X\bm X \right)\left(\cos\left(
  At \right)-i\sin\left( At \right)\bm Y\bm Y \right)\\
&=\cos^2(At)+\sin^2(At)\bm Z\bm Z-i\sin(At)\cos(At)(\bm X\bm X+\bm
  Y\bm Y).
\end{align}
Hence for $At=\pi/4$, we have the $i{\rm SWAP}$ gate
\begin{align}
i{\rm SWAP}=\frac{1}{2}\left[ \openone +\bm Z\bm Z-i(\bm X\bm X+\bm Y\bm Y)
\right]=\begin{pmatrix}
1&0&0&0\\
0&0&i&0\\
0&i&0&0\\
0&0&0&1
\end{pmatrix}.
\end{align}
The $\sqrt{i\rm SWAP}$ gate used in the main text is obtained simply
by halving the evolution time. Together with single qubit rotations around an arbitrary
axis, these gate forms a universal gate set. A CNOT gate for example
can be obtained from single-qubit rotations and two $\sqrt{i\rm SWAP}$
gates~\cite{Nielsen-2000a}.

\section{Code folding}
\begin{figure}[ht]
\includegraphics[width=0.7\textwidth]{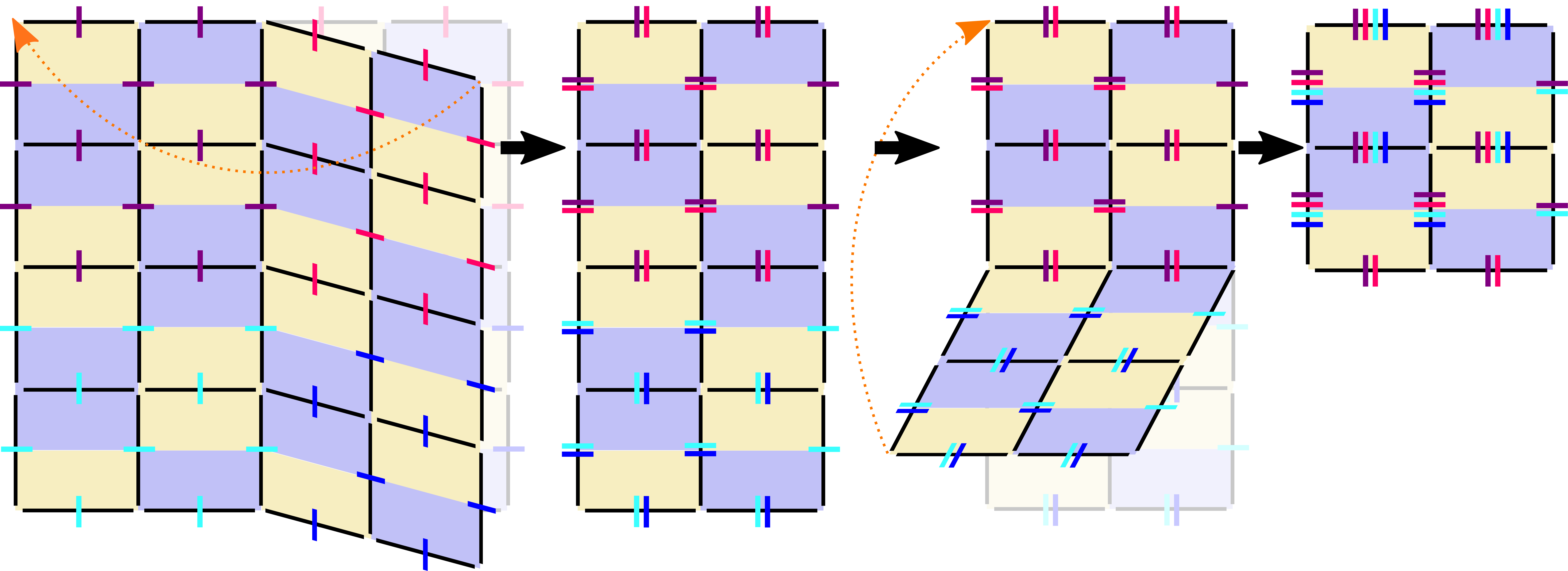}\caption{Code
folding of a $40$ qubit lattice from one qubit per resonator to four and two qubits per resonator.\label{fig:code_folding}}
\end{figure}

The small size of the qubits together with their
tunable ac-field coupling offers the possibility to trade
size with parallel processing capability: Instead of using
one resonator per qubit as described above, each resonator hosts
multiple qubits as depicted schematically in
the rightmost panel of Fig.~\ref{fig:code_folding}. The coupling of each qubit with the ac-field is now controlled by
individual voltage bias lines. In the extreme case where an entire $2^M\times
2^M$ lattice is folded $M$ times onto a single square, all stabilizer
mappings must be made sequentially because no more than one qubit per resonator
can be coupled at a given time. Depending on the experimental
situation however, a partial folding may provide the optimal compromise
between size and speed.

\section{Numerical simulations}
This section provides further numerical results to complement those
presented in the main text.
\subsection{Single-qubit rotation}
Fig.~\ref{fig:single_qubit_gate} illustrates the time evolution of a $2\times 2$ lattice during the
operation of a $\pi$ rotation around the $x$ axis of the qubit at
lattice coordinate $(0,0)$. The same parameters are used as for Fig.~3
(c) of the main text. In particular the field applied to resonator
at $(0, 0)$ is $E_z=0.8\,{\rm V/\mu m}$. We can observe directly the small
resonator population induced in the driven resonator as well as the
effect of the drive on the adjacent resonators. The gate fidelity here
is $98.7\%$.
\begin{figure}[ht]
\includegraphics[width=0.9\textwidth]{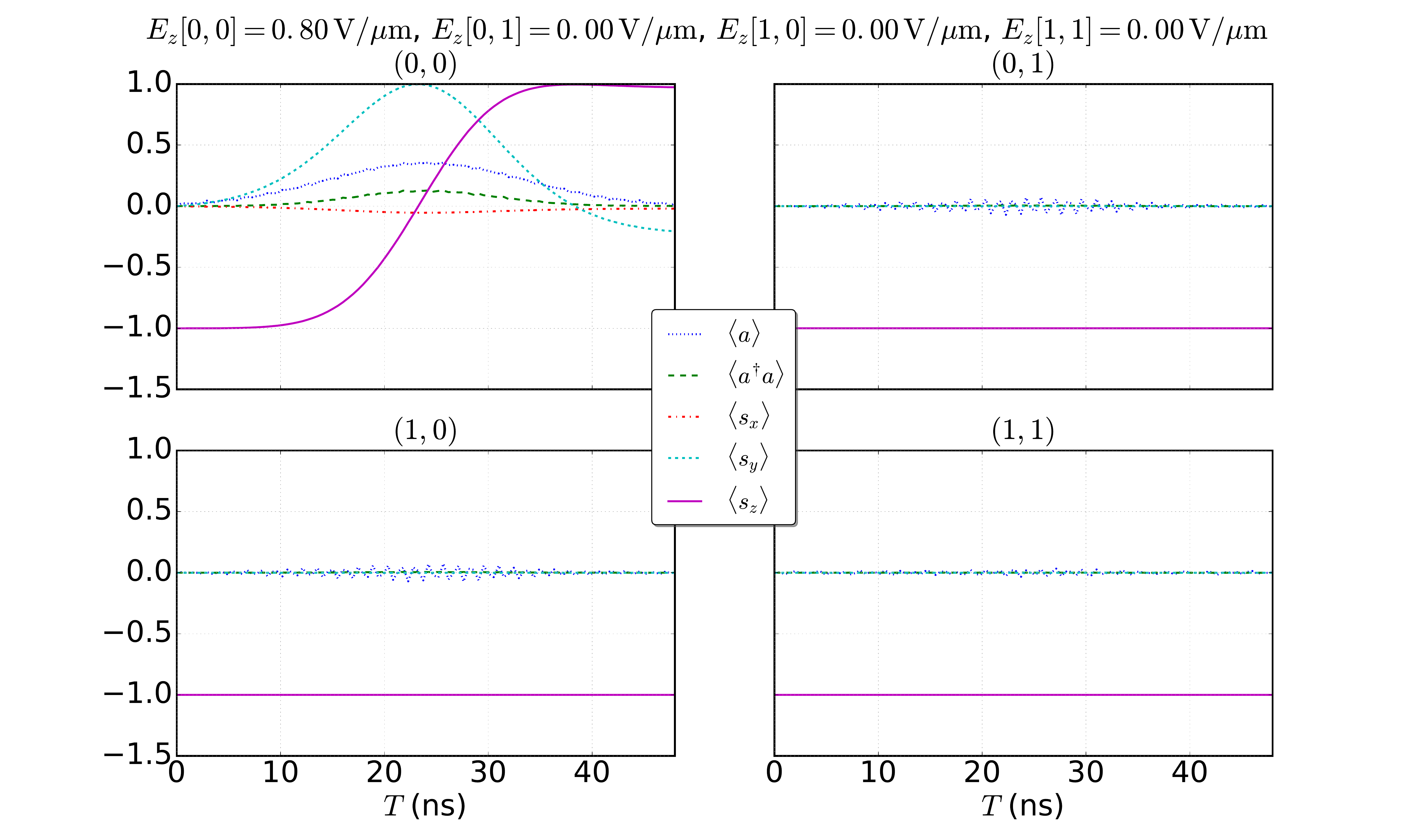}\caption{Time
  evolution in the rotating frame
of a single-qubit rotation $R_x(\pi)$. The initial state is
$\ket{gggg}$ in clock-wise ordering and only the first qubit is
coupled to its cavity. The resonator population due to
resonator-resonator coupling decreases the further the resonator is
from the driven resonator.\label{fig:single_qubit_gate}}
\end{figure}

\subsection{Two-qubit $\sqrt{i\rm SWAP}$ gate}
Fig.~\ref{fig:two_qubit_gate} illustrates the time evolution of a $2\times 2$ lattice during the
operation of a $\sqrt{i\rm SWAP}$ gate between qubits at lattice sites
$(0, 0)$ and $(0,1)$. The same parameters are used as in Fig.~3 (d) of the
main text. A field of strength $E_z=0.9$ is applied to both
resonators at $(0,0)$ and $(0,1)$. The gate fidelity here is $98.1\%$.
\begin{figure}[ht]
\includegraphics[width=0.9\textwidth]{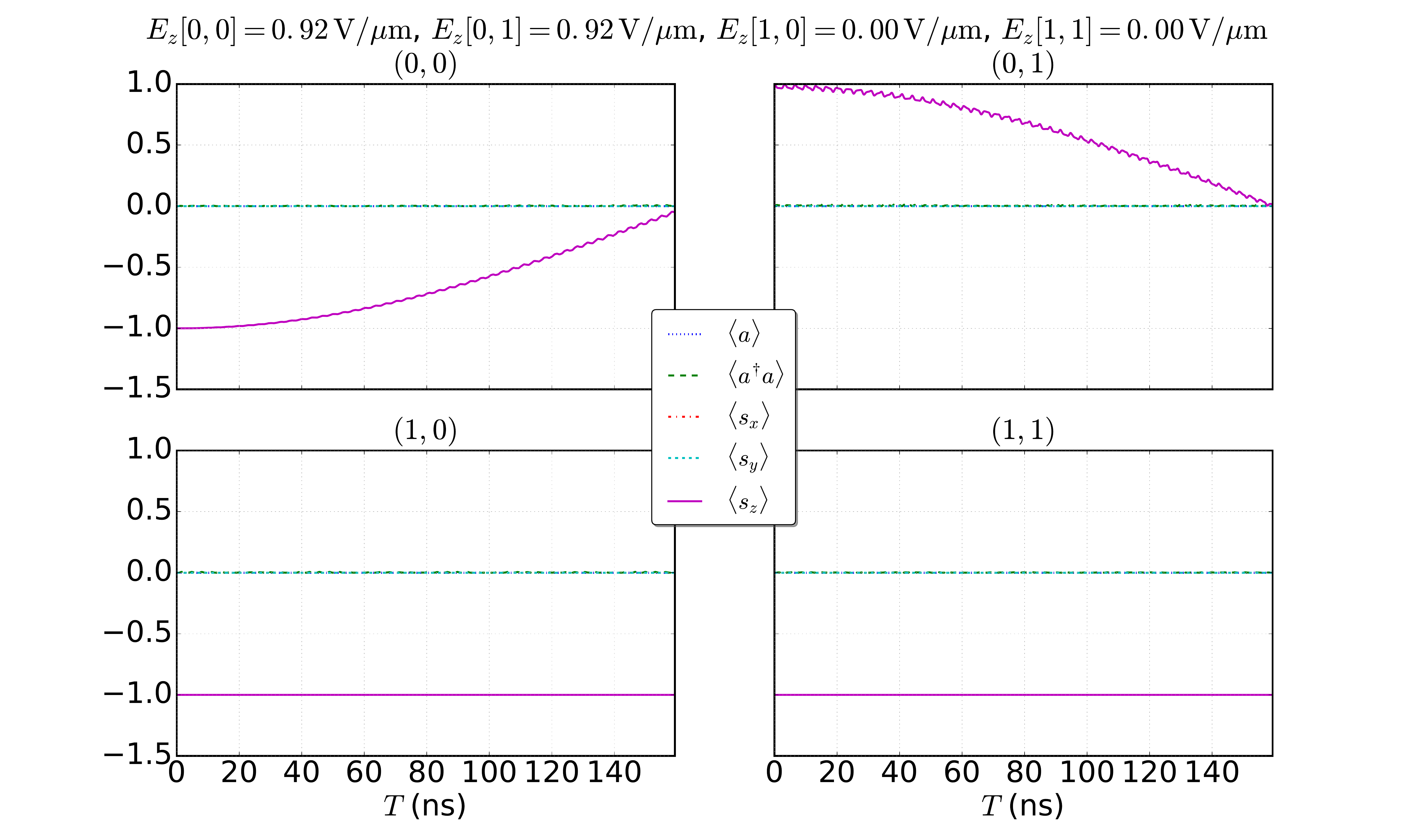}\caption{Time
  evolution
of a two-qubit $\sqrt{i\rm SWAP}$ gate. The initial state is
$\ket{gegg}$ in clock-wise ordering. The last two qubits are not
coupled to their cavities and are seen to be unaffected by
the gate.\label{fig:two_qubit_gate}}
\end{figure}
\end{widetext}
\end{document}